\documentclass[aps,pre,twocolumn,superscriptaddress,showpacs,amsmath,amssymb,footinbib,longbibliography]{revtex4-1}
\usepackage[english]{babel}
\usepackage{amssymb,amsbsy,amsmath}
\usepackage{graphicx}
\usepackage{dcolumn}
\usepackage{bm}
\usepackage{subfigure}
\usepackage{color}
\usepackage{textcomp}
\usepackage[colorlinks,bookmarks=false,citecolor=blue,linkcolor=red,urlcolor=blue]{hyperref}
\newcommand {\subhm} {$\mathcal{H}\left(M\right)$}

\newcommand{\op}[1]{%
    \fontdimen12\textfont3=2pt\fontdimen12\scriptfont3=1.4pt%
    \!\null\mathop{\vphantom{#1}\smash{#1}}\limits_{\sim}\null\!}
\newcommand{\xref}[1]{\protect\ref{#1}}
\newcommand{\figref}[1]{Fig.~\protect\ref{#1}}
\newcommand{\fmref}[1]{(\protect\ref{#1})}
\def\bra#1{\langle \, {#1} \, | \,}
\def\ket#1{\, | \, {#1} \, \rangle}
\newcommand{\braket}[2]{\langle \, {#1} \, | \, {#2} \, \rangle}

\newcommand{\kagome}{kagom\'e}

\renewcommand{\eqref}[1]{Eq.~(\protect\ref{#1})}
\newcommand{\dint}{\text{d}}
\begin{document}
\title{Accuracy of the finite-temperature Lanczos method
  compared to simple typicality-based estimates}

\author{J\"urgen Schnack}
\email{jschnack@uni-bielefeld.de}
\affiliation{Fakult\"at f\"ur Physik, Universit\"at Bielefeld, Postfach 100131, D-33501 Bielefeld, Germany}
\author{Johannes Richter}
\email{Johannes.Richter@physik.uni-magdeburg.de}
\affiliation{Institut f\"ur Theoretische Physik, Universit\"at Magdeburg, P.O. Box 4120, D-39016 Magdeburg, Germany}
\affiliation{Max-Planck-Institut f\"{u}r Physik Komplexer Systeme,
        N\"{o}thnitzer Stra{\ss}e 38, 01187 Dresden, Germany}
\author{Robin Steinigeweg}
\email{rsteinig@uos.de}
\affiliation{Fachbereich Physik, Universit\"at
  Osnabr\"uck, Barbarastr. 7, D-49076 Osnabr\"uck, Germany}

\date{\today}

\begin{abstract}
We study trace estimators for equilibrium thermodynamic
observables that rely on the idea of typicality and
derivatives thereof such as the finite-temperature Lanczos
method (FTLM). As numerical examples quantum spin systems are
studied. Our initial aim was to identify pathological examples
or circumstances, such as strong frustration or unusual
densities of states, where these methods could fail. It turned
out, that all investigated systems allow such 
approximations. Only at temperatures of the order of the
lowest energy gap the convergence is somewhat slower in the number of
random vectors over which observables are averaged.
\end{abstract}


\maketitle

\section{Introduction}

One way to approximate thermodynamic quantities is to rely on \emph{trace 
estimators}. These schemes became very popular in recent years since they turn 
out to be rather (or astonishingly) accurate and in addition a valuable 
alternative in cases where Quantum Monte Carlo suffers from the sign problem. 
Trace estimators approximate a trace by a simple evaluation of an expectation 
value with respect to a random vector
\cite{Ski:88,Hut:CSSC89,DrS:PRL93,JaP:PRB94,SiR:IJMPC94,GoM:Stanford97,WWA:RMP06,AvT:ACM11,RoA:FCM15,SAI:NM17},
i.e.,
\begin{eqnarray}
\label{E-1-1}
\text{tr}
\left(\op{O}\right)
&\approx&
\bra{r}\op{O}\ket{r}
\;
\frac{\text{dim}({\mathcal H})}{\braket{r}{r}}
\ .
\end{eqnarray}
Here $\op{O}$ is the operator of interest, and $\ket{r}$ is a 
vector (pure state) drawn at random from a high-dimensional
Hilbert space. The factor on the r.h.s. takes care of the
dimension. If not mentioned otherwise the vector $\ket{r}$ will be
normalized. 

The complex components $r_{\nu}$ of the vector $\ket{r}$ 
with respect to a chosen orthonormal basis $\{\ket{\nu}\}$,
\begin{eqnarray} \label{eq:te}
\label{E-1-2}
\ket{r}
&=&
\sum_{\nu}\;
r_{\nu} \ket{\nu}
\ ,
\end{eqnarray}
are supposed to follow a Gaussian distribution with zero
mean (Haar measure \cite{CoS:CMP06,BaG:PRL09,Rei:NC16}). Under
unitary basis transforms this  
distribution remains Gaussian, i.e, also in the energy
eigenbasis.

The latter fact is the essential difference to the eigenstate
thermalization hypothesis (ETH)
\cite{Deu:PRA91,Sre:PRE94,RDO:N08}, which assumes that the
approximation in \eqref{E-1-1} 
also holds if $\ket{r}$ is replaced by an individual energy
eigenstate (with the trace operation being performed in a
microcanonical energy shell). While it is well 
established that the ETH is indeed satisfied for few-body observables in 
generic nonintegrable models \cite{AKP:AP16,BIS:PR16},
counterexamples are also known to exist, with 
integrable models \cite{AKP:AP16} and many-body localized
systems \cite{NaH:ARCMP15,AAB:RMP19} as the most prominent 
ones. However, this discussion does not interfere with the
investigations presented in our paper simply for the reason that
individual energy eigenstates are obviously no Gaussian random
superpositions of energy eigenstates, a prerequisite we require
for \fmref{E-1-1} to work.

The traces to be delt with in this article appear in equilibrium
statistical physics where they are evaluated for static
operators such as $\exp\{-\beta\op{H}\}$ and
$\op{O}\exp\{-\beta\op{H}\}$ yielding the partition function and thermal 
expectation values of observables $\op{O}$, respectively. In this context, 
different schemes relying on trace estimators, sometimes termed 
\emph{typicality} or \emph{(microcanonical) thermal pure 
quantum states} \cite{IMN:IEEE19,SuS:PRL12,SuS:PRL13,OAD:PRE18},
have been 
very successfully employed in the field of correlated electron systems to 
evaluate magnetic observables, see
e.g. \cite{ABK:PRB75,DRdV:ZP89,dVDR:PRB93,JaP:PRB94,Dag:RMP94,JaP:AP00,ADE:PRB03,ZST:PRB06,ScW:EPJB10,PrB:SSSSS13,USL:JMMM13,HaS:EPJB14,PHK:PRB14,SGB:PRB15,SHP:PRB16,ScT:PR17,OAD:PRE18,SSR:PRB18,PrK:PRB18,DKR:PRB18,RiS:PRB19,RKK:PRB19},
but also elsewhere \cite{MHL:CPL01,HLM:JCP02}. 
Although some estimates for the accuracy of such schemes have
been provided analytically
\cite{Hut:CSSC89,AvT:ACM11,RoA:FCM15,SAI:NM17,PRE:COR17,HaD:PRE00,IiE:PRL03} as well
as numerically \cite{ScW:EPJB10,LSY:A13,RoA:FCM15,OAD:PRE18,SSR:PRB18}, more confidence 
into
the approximation seems to be desirable in particular in view of
some scepticism \cite{WWA:RMP06,WCW:PRR19}.

We therefore present large scale numerical calculations for 
bipartite and geometrically frustrated archetypical spin systems, together with 
a detailed analysis of the statistical errors. The use of
conserved quantities (good quantum numbers) and a related decomposition of the 
full Hilbert space according to irreducible representations is as 
well addressed. We find that the gross estimation of the relative variance of 
an observable $\op{O}$ in leading order of system properties 
(thermally occupied levels),
\begin{eqnarray}
\label{E-1-3}
{\delta \langle\op{O}\rangle}
&\simeq&
{\langle\op{O}\rangle}\frac{\alpha}{\sqrt{Z_{\text{eff}}}}
\ ,\qquad
Z_{\text{eff}} = \text{tr}\left(e^{-\beta (\op{H}-E_0)}\right)
\ ,
\end{eqnarray}
with $E_0$ being the ground state energy, is indeed roughly fulfilled for 
$\alpha=1$ and not too low temperatures, compare \cite{JaP:AP00,HaD:PRE00,PRE:COR17}.
While \eqref{E-1-3} is 
found to hold approximately, it is worth pointing out that additionally
an upper bound for ${\delta \langle\op{O}\rangle}$ can be
derived, see e.g.~\cite{SuS:PRL12}. This upper bound does not
only justify trace estimators for  
static quantities but also for dynamic quantities \cite{BaG:PRL09,MoS:JPSJ14} 
such as time-dependent correlation functions
\cite{IiE:PRE04,SGB:PRL14,SGB:PRB15,IiE:PRL03,ElF:PRL13}. The concept  
of \emph{dynamical typicality} further plays a central role for the foundations 
of statistical mechanics and thermodynamics
\cite{GMM:09,BRG:18,ReG:PA19}. In this paper, however, we 
focus on static observables such as heat capacity and
susceptibility. For such quantities one can also show that ${\delta
  \langle\op{O}\rangle}=0$ for $T=0$ for systems with
non-degenerate ground states \cite{JaP:AP00,PRE:COR17}. 

For a deeper understanding of the actual numerical method 
applied later in this paper it is helpful to note that two rather 
different approximations are employed for the evaluation of thermal 
averages. The first approximation is provided by the trace estimator 
in \fmref{E-1-1}. The second approximation is 
provided by the Krylov space expansion of the exponential which yields a 
spectral representation that covers the true spectrum in a coarse grained manner 
\cite{SSR:PRB18}. These issues will be addressed in the following. 
Note that the second approximation might be replaced by other 
approaches which solve the imaginary-time Schr\"odinger equation iteratively 
such as fourth-order Runge-Kutta
\cite{SGB:PRL14,SGB:PRB15,ElF:PRL13} or Chebyshev polynomials 
\cite{WWA:RMP06}. 

The paper is organized as follows. In Section \ref{sec-2} we recapitulate 
typicallity-based estimators. In Section~\ref{sec-3} we present our numerical 
examples both for frustrated and unfrustrated spin systems. The article closes
with a discussion in Section~\ref{sec-4}.

\section{Method}
\label{sec-2}

There are many sound motivations and derivations of the idea that traces can be 
accurately approximated by expectation values with respect to 
several or single random vectores
\cite{Hut:CSSC89,AvT:ACM11,RoA:FCM15,SAI:NM17} as in
\fmref{E-1-1}, that we do not want to repeat it here. 
We want to motivate the older idea of \emph{trace estimators}
using the more recent concept of \emph{typicality}.
The idea of typicality in the context of trace estimators for
physical quantities such as the partition function means that
the overwhelming majority of all random vectors that one can
draw in a high-dimensional Hilbert space consists of virtually
equivalent vectors and corresponds 
to a situation of infinite temperature, where all expansion
coefficients of the density matrix with respect to an 
orthonormal basis would be just
about the same. In one of the very first realizations by
Hutchinson \cite{Hut:CSSC89} this was explicitely built into the
method by generating random vectors with unit entries but random
sign (Rademacher random vectors). Later it turned out 
that unbiased estimators can also be set up with other
distributions, as for instance Gaussian distributions
\cite{SAI:NM17} (cf. Haar measure
\cite{CoS:CMP06,BaG:PRL09,Rei:NC16}). 

In the ideal case one could compute a typicallity-based
expectation value, depending on temperature $T$ and magnetic field $B$,
using just one single random state $\ket{r}$,
i.e. 
\begin{eqnarray}
\label{E-2-A}
O^{\text{r}}(T,B)
&\approx&
\frac{\bra{r}\op{O}e^{-\beta \op{H}}\ket{r}}
     {\bra{r}e^{-\beta \op{H}}\ket{r}}
\ .
\end{eqnarray}
Numerical examples indicate that one single random state
indeed works well for dense spectra and large enough Hilbert
spaces \cite{SSR:PRB18}, where the notion of ``large'' clearly 
depends on temperature. For high temperatures, already
dimensions of oder $10^3$ 
can be sufficient, while for low temperatures, the required
dimension increases substantially, cf.~\fmref{E-1-3}.

One could naively assume that this approximation could be
improved by the following mean with respect to a set of $R$
different random states, 
\begin{eqnarray}
\label{E-2-B}
O^{\text{mean}}(T,B)
&\approx&
\frac{1}{R}
\sum_{r=1}^R\;
\frac{\bra{r}\op{O}e^{-\beta \op{H}}\ket{r}}
     {\bra{r}e^{-\beta \op{H}}\ket{r}}
\ ,
\end{eqnarray}
but this is not true, as we will see in the next section.

It is instead more accurate to improve the involved traces 
separately, albeit with respect to the same set of 
random vectors in numerator and denominator,
\begin{eqnarray}
\label{E-2-C}
O^{\text{FTLM}}(T,B)
&\approx&
\frac{\sum_{r=1}^R\;\bra{r}\op{O}e^{-\beta \op{H}}\ket{r}}
     {\sum_{r=1}^R\;\bra{r}e^{-\beta \op{H}}\ket{r}}
\ .
\end{eqnarray}
The latter corresponds to the scheme that is used in the
finite-temperature Lanczos method (FTLM)
\cite{JaP:PRB94,JaP:AP00,PRE:COR17}.

Technical details particularly concern the evaluation of 
${\bra{r}e^{-\beta \op{H}}\ket{r}}$, i.e.\ the application of the
exponential. FTLM employs a Krylov space expansion, i.e.\ a
spectral representation of the exponential in a Krylov space
grown from $\ket{r}$ as starting vector. A similar idea can be
realized using Chebyshev polynomials \cite{WWA:RMP06}.

In addition, if the Hamiltonian $\op{H}$ possesses symmetries, these can be 
employed by decomposing the full Hilbert space into
mutually orthogonal subspaces according to the irreducible
representations of the employed symmetry
\cite{ScW:EPJB10,HaS:EPJB14} leading for the partition function
to  
\begin{eqnarray}
\label{E-2-D}
Z^{\text{FTLM}}(T,B)
&\approx&
\sum_{\gamma=1}^\Gamma\;
\frac{\text{dim}({\mathcal H}(\gamma))}{R}
\nonumber
\\
&&
\times
\sum_{r=1}^R\;
\sum_{n=1}^{N_L}\;
e^{-\beta \epsilon_n^{(r)}} |\braket{n(r)}{r}|^2
\ .
\end{eqnarray}
${\mathcal H}(\gamma)$ denotes the subspace that belongs to 
the irreducible representation $\gamma$, $N_L$ is the dimension
of the generated Krylov space, and $\ket{n(r)}$ is the n-th
eigenvector of $\op{H}$ in this Krylov space with seed $\ket{r}$
and energy eigenvalue $\epsilon_n^{(r)}$. $N_L$ is chosen such,
that the ground state energy in the respective subspace is converged
to numerical accuracy. For large subspaces this requires $N_L$ of
the order of $300\dots 500$. 
The method is publicly available with the program
\verb§spinpack§ \cite{spin:256,RiS:EPJB10}.

\begin{figure}[ht!]
\centering
\includegraphics*[clip,width=0.85\columnwidth]{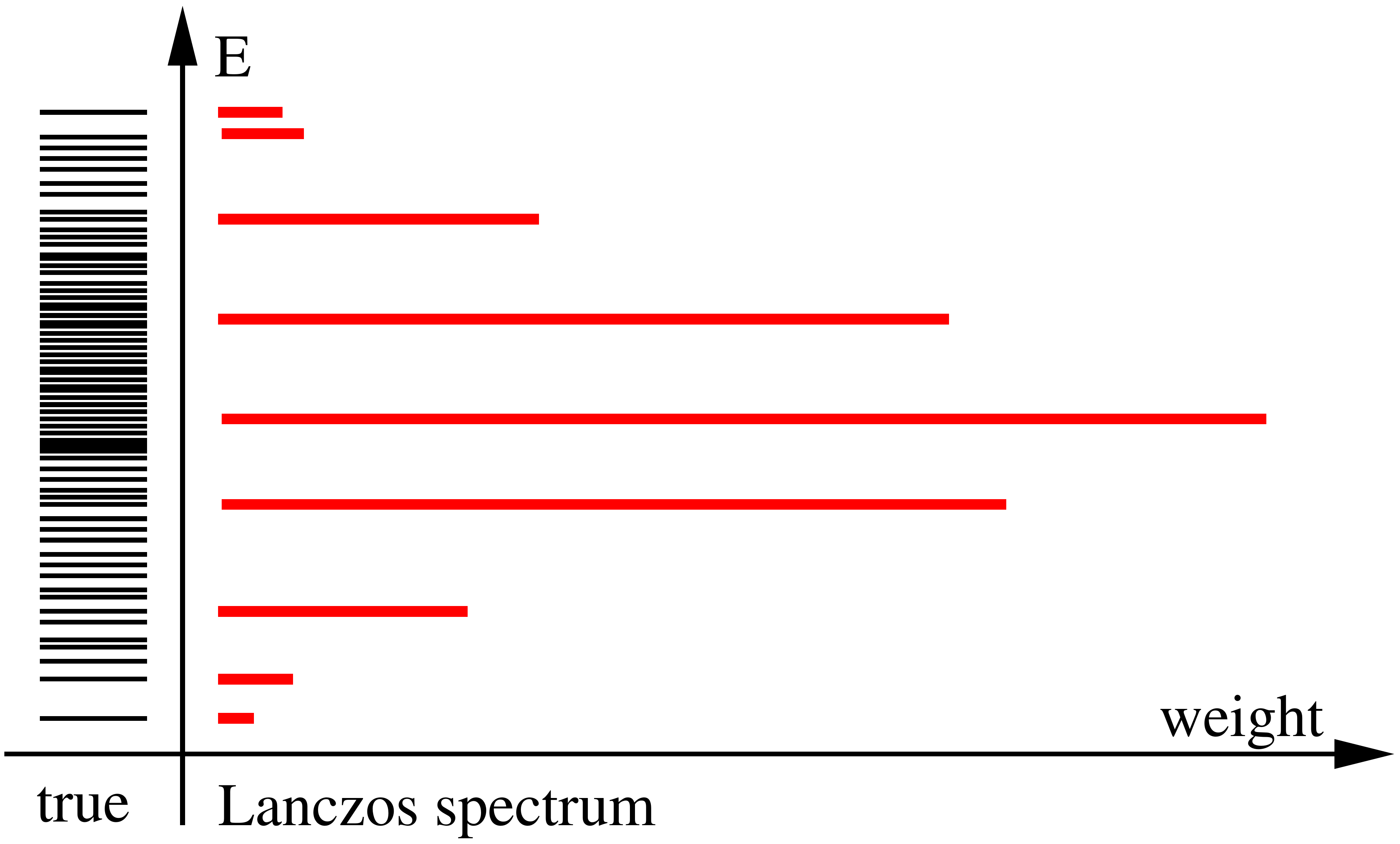}
\caption{Schematic representation of the approximation of a true
  energy spectrum by Lanczos energies $\epsilon_n$ and their
  weights $w_n(r)=|\braket{n(r)}{r}|^2$, compare Eq.\ \fmref{E-2-D}.} 
\label{ftlm-accuracy-f-ls}
\end{figure}

Effectively, the method provides a coarse grained coverage of
the density of states by means of energy representatives
$\epsilon_n^{(r)}$ that come together with weight factors 
$w_n(r)=|\braket{n(r)}{r}|^2$, thus emulating the true density
of levels in the vicinity of the energy representative. 
In this respect the method has got some similarities to the
classical Wang-Landau sampling \cite{WaL:PRL01,ZST:PRL06,SBL:PRB07}.

Approximations of the type discussed here essentially suffer from
two sources of error: (i) the choice of random vectors
in a Hilbert space of large but finite dimension and (ii) the expansion of the 
exponential in the respective Krylov spaces grown from the random vectors. 
The error due to choosing a \emph{single} random vector is upper bounded by
${\cal O}(1/\sqrt{Z_\text{eff}})$. In addition, one can prove minimal
numbers $R$ of random vectors if a certain minimal probability 
$(1-\delta)$ is required to stay below a certain relative
deviation $\varepsilon$ \cite{RoA:FCM15} 
\begin{eqnarray}
\label{E-2-Z}
\text{Pr}
\left(
|\text{tr}_D^R(\op{A})-\text{tr}(\op{A})|
\leq \varepsilon\ \text{tr}(\op{A})
\right)
\geq
1-\delta
\ .
\end{eqnarray}
Here $\text{tr}_D^R(\op{A})$ denotes a trace with respect to a
mathematical distribution $D$ of random numbers, that is
averaged over $R$ realizations.
For the Hutchinson method one obtains
$R\geq 6 \varepsilon^{-2}\text{ln}(2/\delta)$, for a Gaussian distribution 
$R\geq 8 \varepsilon^{-2}\text{ln}(2/\delta)$ \cite{RoA:FCM15}.

The error related to the specific approximation of the
exponential, e.g. Krylov or Chebyshev, is not so simple to
quantify, which is one of the reasons for our numerical
study. The central question is how accurate the coarse grained
Lanczos spectrum represents the true spectrum in the partition
function for various temperatures. It is evident from the
schematic representation given in \figref{ftlm-accuracy-f-ls} that the
Lanczos spectrum does not provide  
accurate resonance frequencies for more than the lowest energy
gaps, even if these are indeed very accurate thanks to the (in $N_L$)
exponentially fast convergence 
of extremal eigenvalues in the Lanczos method \cite{Saa:SIAMJNA80}.
In addition, symmetries of the Hamiltonian can not only be used to yield
smaller orthogonal subspaces ${\mathcal H}(\gamma)$ according to
\eqref{E-2-D} which helps to access larger system sizes and/or to make 
calculations faster and less memory consuming, but also to generate a larger
number of very accurate Lanczos energy values since they are extremal in their 
respective subspaces. However, even if the lowest Lanczos energies are very 
accurate, this does not automatically imply that the related weights $w_n(r)$ 
share the same superb accuracy. The low-temperature behavior thus remains to be
further explored. Only at $T=0$ the observables are bound to be
correct as long as the the random vector has got some overlap
with the true (non-degenerate) ground state
\cite{PRE:COR17,BRG:18}. One the other hand, 
for high temperatures one can show that  
the expansion of the exponential in Krylov space up to
order $N_L$ corresponds to a high-temperature series expansion
of equilibrium expectation values up to the same order
\cite{PRE:COR17}. Overall, relation \fmref{E-1-3} could be
derived in Ref.~\cite{PRE:COR17} under rather reasonable 
assumptions. This relation will be our reference in the upcoming
part of the paper.

In the following we estimate the uncertainty of a physical quantity
approximated by a trace estimator by repeating the numerical evaluation
$N_S$ times. The generated set of results is considered as a
statistical sample, for which we define the standard
deviation of the observable in the following way:
\begin{eqnarray}
\label{E-2-E}
\delta(O)
&=&
\sqrt{
\frac{1}{N_S}
\sum_{r=1}^{N_S}\;
\left(
O^{\text{m}}(T,B)
\right)^2
-
\left(
\frac{1}{N_S}
\sum_{r=1}^{N_S}\;
O^{\text{m}}(T,B)
\right)^2
}
\nonumber
\\
&=&
\sqrt{
\overline{\left(
O^{\text{m}}(T,B)
\right)^2}
-
\left(\overline{
O^{\text{m}}(T,B)}
\right)^2
}
\ .
\end{eqnarray}
$O^{\text{m}}(T,B)$ is either evaluated according to
\eqref{E-2-A} or to \eqref{E-2-C}, depending on whether the
fluctuations of approximations with respect to one random vector
or with respect to an average over $R$ vectors shall be
investigated (cf. the following examples).

In the following numerical examples two observables are
considered, zero-field susceptibility and heat capacity. Both
are evaluated as variances of magnetization and energy,
respectively, i.e.
\begin{eqnarray}
\label{E-2-24}
\chi(T)
&=&
\frac{(g \mu_B)^2}{k_B T}\,
\Bigg\{
\left\langle(\op{S}^z)^2\right\rangle
-
\left\langle\op{S}^z\right\rangle^2
\Bigg\}
\\
C(T)
&=&
\frac{k_B}{(k_B T)^2}\,
\Bigg\{
\left\langle\op{H}^2\right\rangle
-
\left\langle\op{H}\right\rangle^2
\Bigg\}
\ .
\end{eqnarray}

\section{Numerical results}
\label{sec-3}

In this section we investigate various spin systems.
They are of finite size and modeled by Heisenberg Hamiltonians 
augmented with a Zeeman term, i.e. 
\begin{eqnarray}
\label{E-2-1}
\op{H}
&=&
-2 \;
\sum_{i<j}\;
J_{ij}
\op{\vec{s}}_i \cdot \op{\vec{s}}_j
+
g \mu_B\, B\,
\sum_{i}\;
\op{s}^z_i
\ ,
\end{eqnarray}
where the first sum runs over ordered pairs of spins (``$-2J$''
convention used). Our
original intention was to identify systems and circumstances
where the approach of Eq.~\fmref{E-2-C} fails. But none of the
investigated systems turned out to be (systematically)
intractable.

\begin{figure}[ht!]
\centering
\includegraphics*[clip,width=0.69\columnwidth]{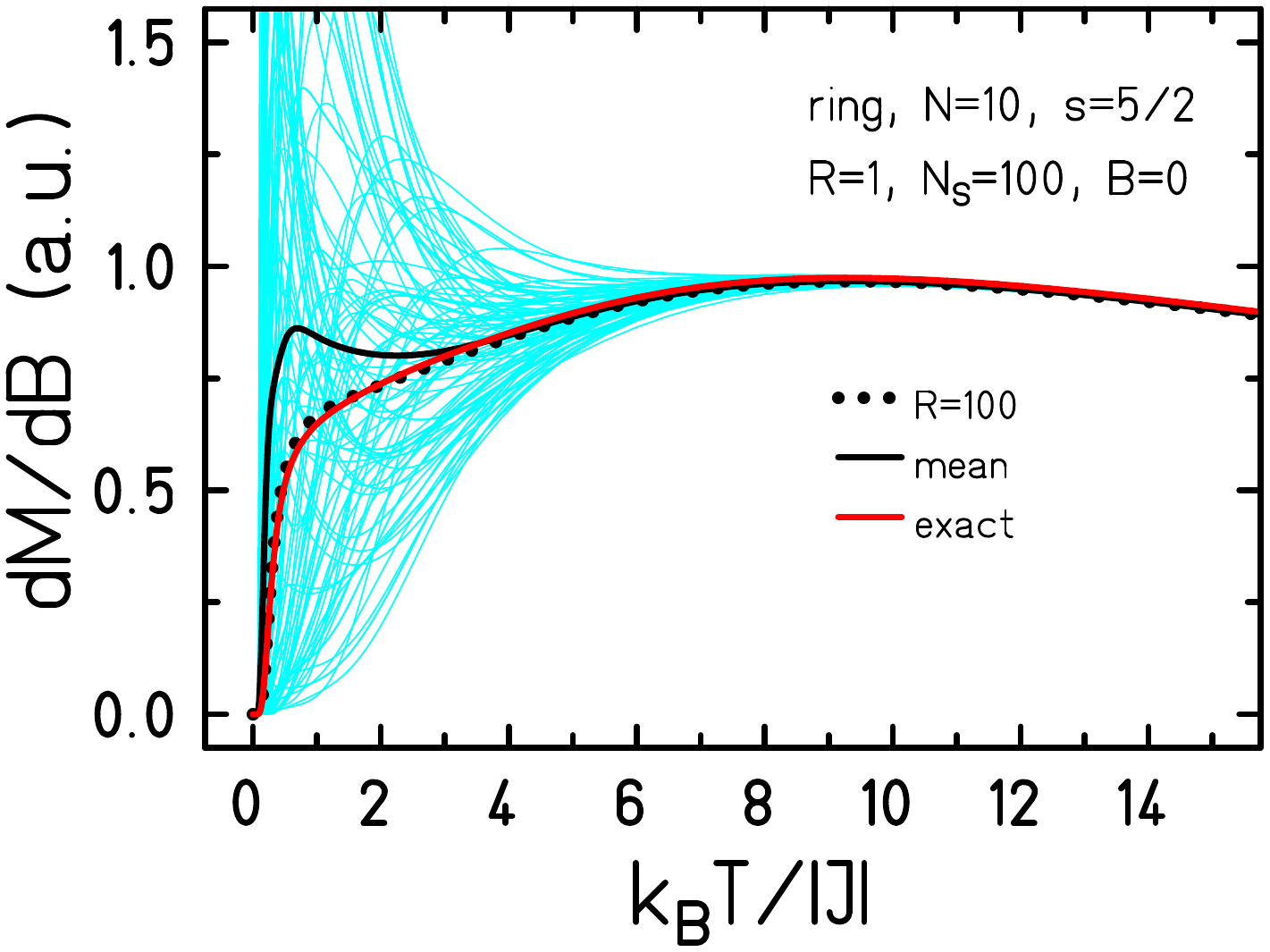}
\includegraphics*[clip,width=0.69\columnwidth]{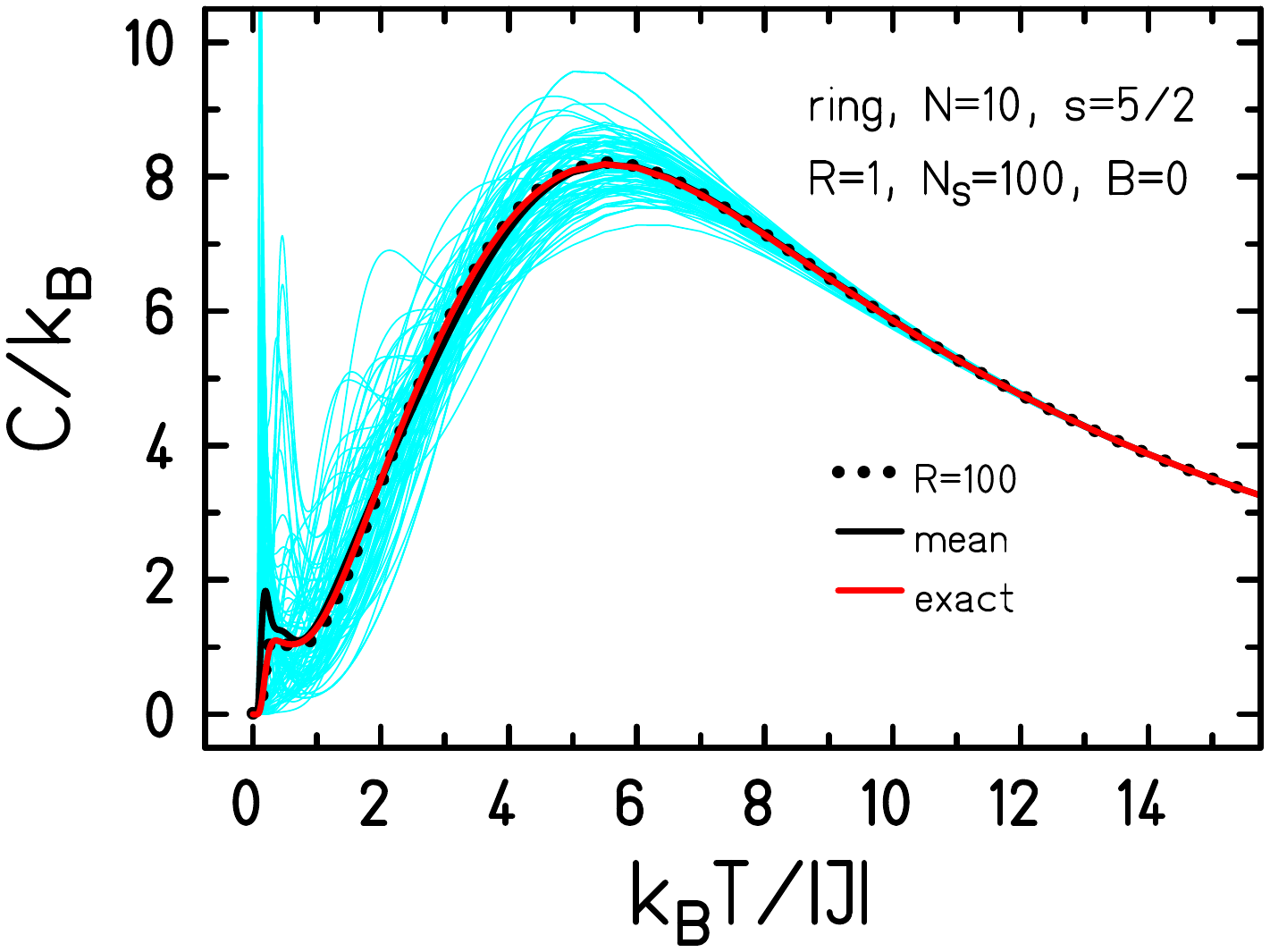}
\caption{Spin ring $N=10$, $s=5/2$: The
  light-blue curves depict 100 different estimates of the
  differential susceptibility as well as the heat capacity using
  single vectors.
  Mean values as well as the exact result are also
  presented.} 
\label{ftlm-accuracy-f-A1}
\end{figure}

\begin{figure}[ht!]
\centering
\includegraphics*[clip,width=0.69\columnwidth]{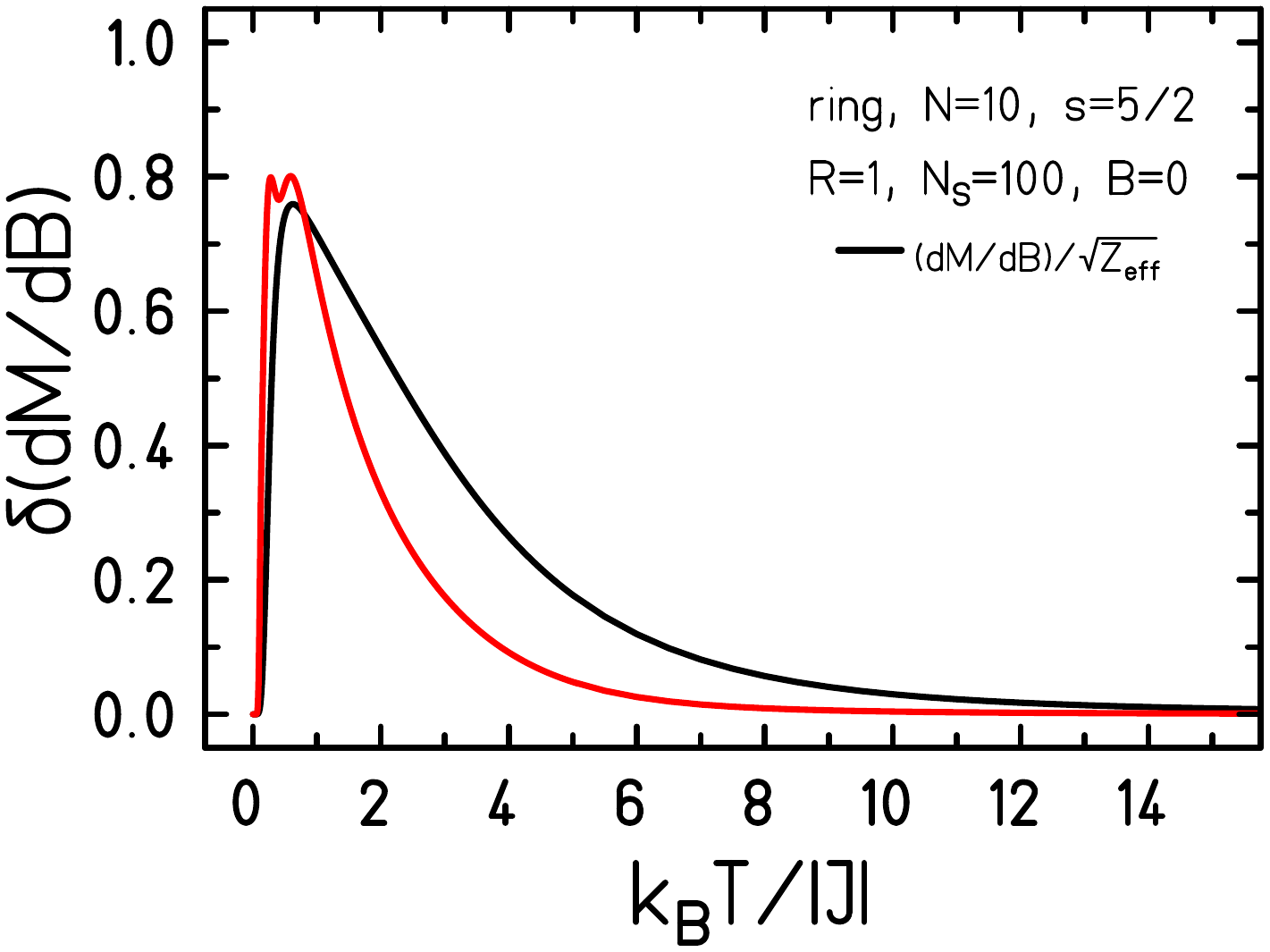}
\includegraphics*[clip,width=0.69\columnwidth]{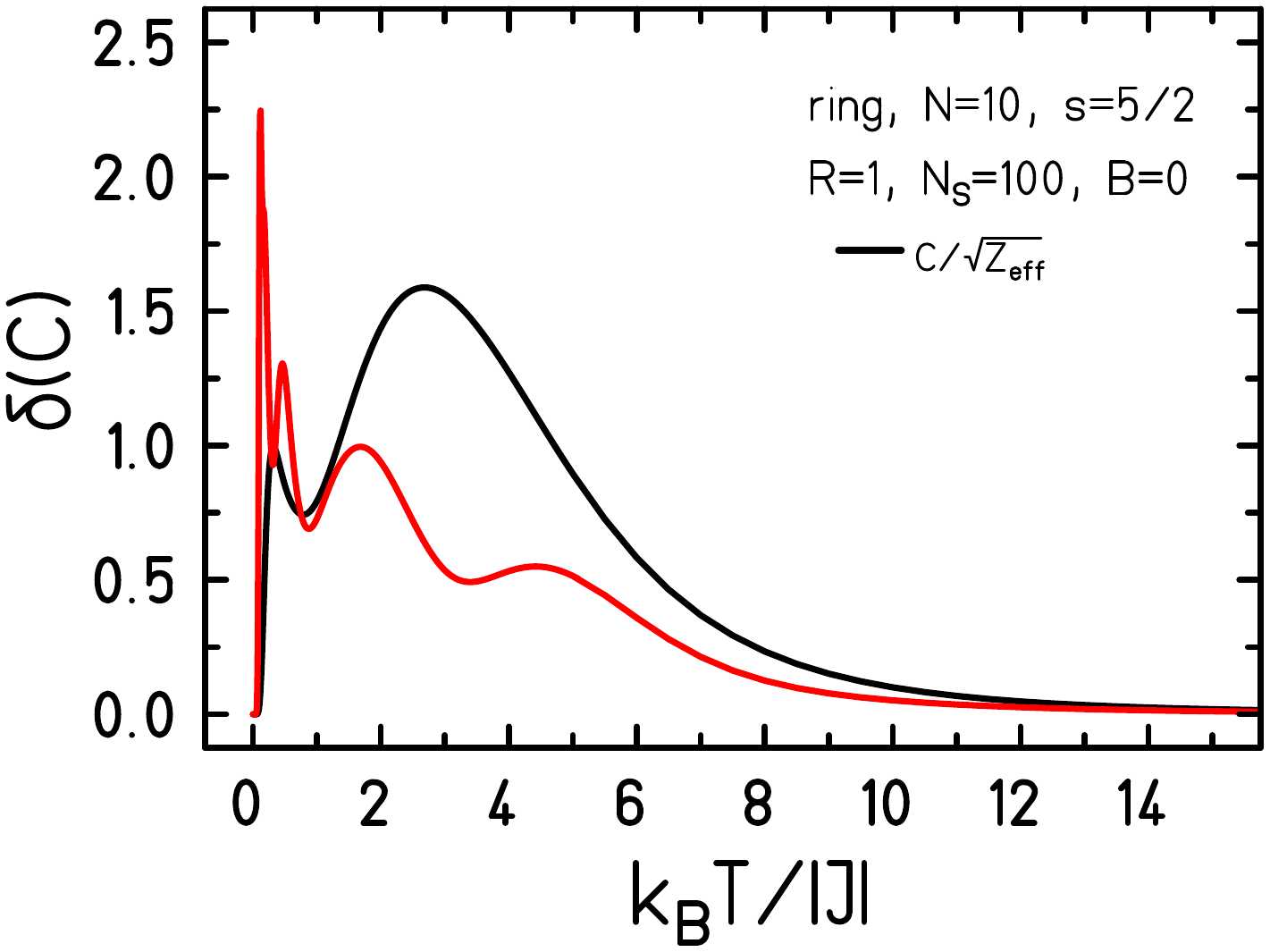}
\caption{Spin ring $N=10$, $s=5/2$: Standard
  deviation (red) of the differential 
  susceptibility as well as the heat capacity compared to the
  error estimate (black).} 
\label{ftlm-accuracy-f-A2}
\end{figure}

\subsection{Spin ring}
\label{sec-3-1}

As a first example we examine a spin ring with $N=10$ spins
$s=5/2$ and nearest-neighbor antiferromagnetic interaction.
This system exists as a magnetic molecule (abbreviated
Fe$_{10}$), and it is called a ``ferric wheel''
\cite{TDP:JACS94}. Although the dimension of the total Hilbert
space is 60,466,176, the Heisenberg Hamiltonian can be
diagonalized completely thanks to the high symmetry (SU(2) and
C$_{10}$) \cite{ScS:IRPC10}.

Figure~\xref{ftlm-accuracy-f-A1} shows $N_S=100$ calculations
of the differential susceptibility as well as the heat capacity 
according to Eq.\ \fmref{E-2-A}, i.e.\ using a single random vector
for each estimate together with the means according to
Eqs.~\fmref{E-2-B} and \fmref{E-2-C}. For the calculation 
$\op{S}^z$-symmetry was employed. In addition the exact result
is depicted. One can clearly see that the estimates using a
single random vector fluctuate largely for temperatures less
than ten times the coupling. Nevertheless, if the estimates are
joined in an FTLM fashion according to \eqref{E-2-C}, the result
for $R=100$ can be hardly distinguished from the exact
calculation. A simple mean according to \eqref{E-2-B} fails.

A statistical analysis of the set of estimates
$O^{\text{r}}(T,B)$ in \figref{ftlm-accuracy-f-A2} reveals that 
the error estimate \fmref{E-1-3} with $\alpha=1$ indeed accurately 
describes the standard deviation. Only for
temperatures smaller than the exchange coupling larger
deviations can be observed, but they do not exceed the
error estimate much (e.g. by orders of magnitude).

\subsection{Cuboctahedron \& Icosidodecahedron}
\label{sec-3-2}

As a second example we choose two frustrated polytopes: the
cuboctahedron as well as the icosidodecahedron with
antiferromagnetic nearest-neighbor interactions. Not only do both
exist as magnetic molecules
\cite{MSS:ACIE99,MLS:CPC01,BKH:CC05,TMB:ACIE07,SPK:PRB08,TMB:CC09,PPS:CPC16},
they are also itimately 
related to the \kagome\ lattice antiferromagnet
\cite{RLM:PRB08,SHL:PRE17,Sch:CP19}. In contrast to the bipartite spin
ring discussed above, these spin systems possess a rather dense
spectrum with for instance several to many singlet states below
the first triplet state (a hallmark of geometric frustration).

\begin{figure}[h]
\centering
\includegraphics*[clip,width=0.69\columnwidth]{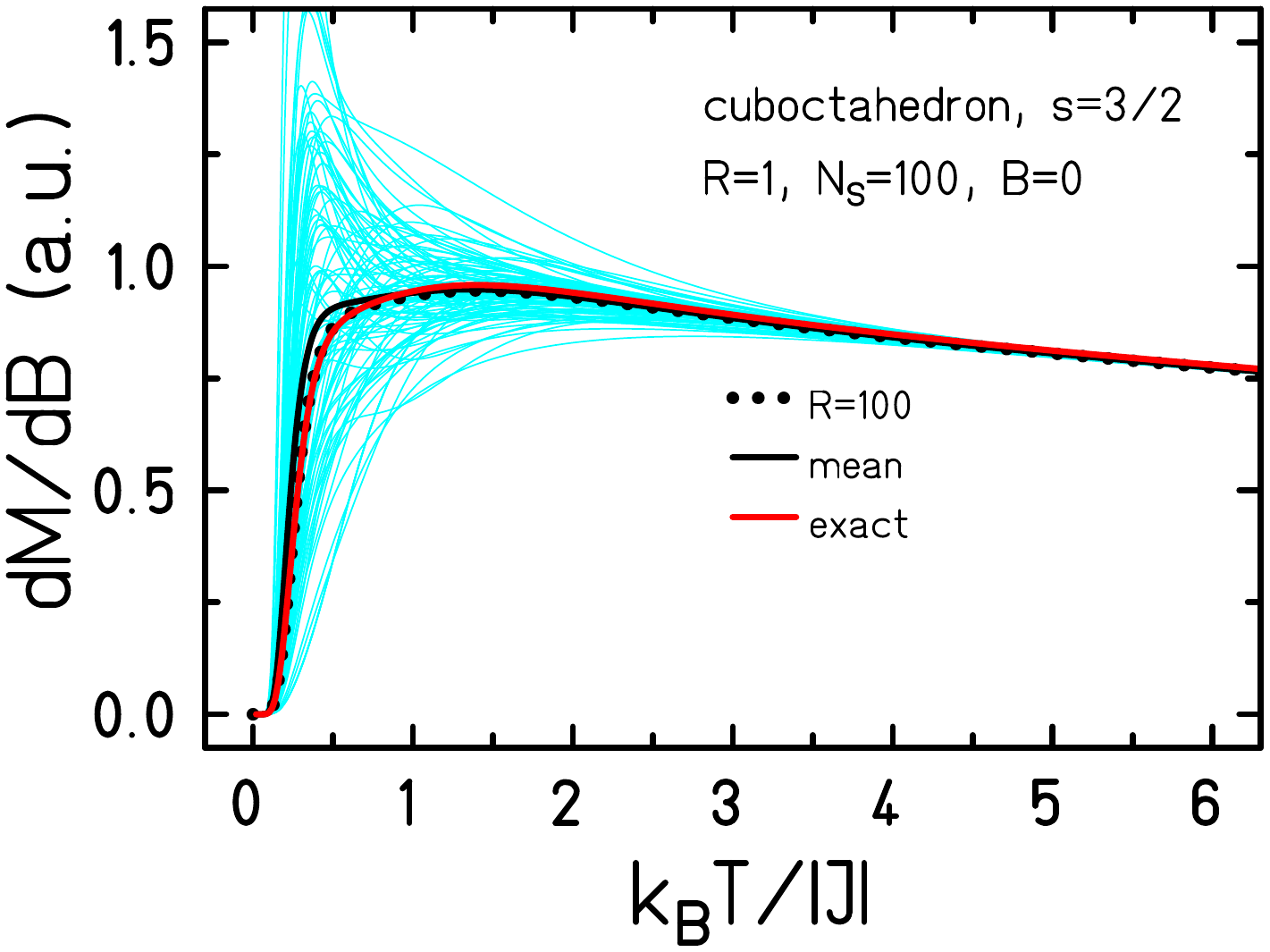}
\includegraphics*[clip,width=0.69\columnwidth]{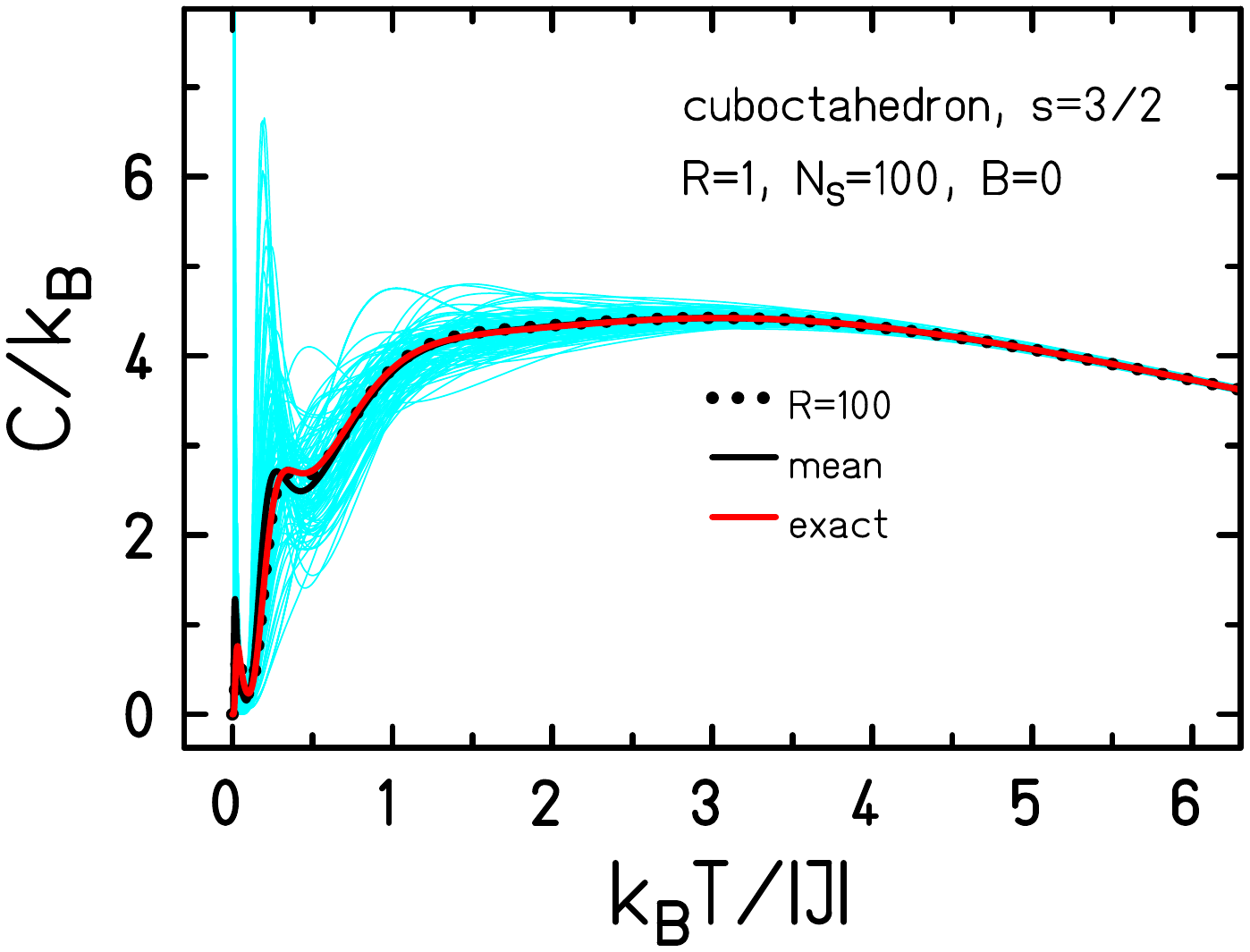}
\caption{Cuboctahedron $N=12$, $s=3/2$: The
  light-blue curves depict 100 different estimates ($R=1$) of the
  differential susceptibility as well as the heat capacity.
  Mean values as well as the exact result are also
  presented.}
\label{ftlm-accuracy-f-B1}
\end{figure}

For the cuboctahedron, that has 12 spin sites, we choose a single-spin
quantum number of $s=3/2$ since we can still completely
diagonalize the Hamiltonian using symmetries although the
dimension of the total Hilbert space is 16.777.216
\cite{ScS:P09}. As can be seen in \figref{ftlm-accuracy-f-B1},
the magnetic observables fluctuate below a temperature of five
times the coupling when evaluated with respect to a single
random vector. Aggregating them into an FTLM estimate with
$R=100$ again yields a very good approximation compared to the
exact result.

\begin{figure}[ht!]
\centering
\includegraphics*[clip,width=0.69\columnwidth]{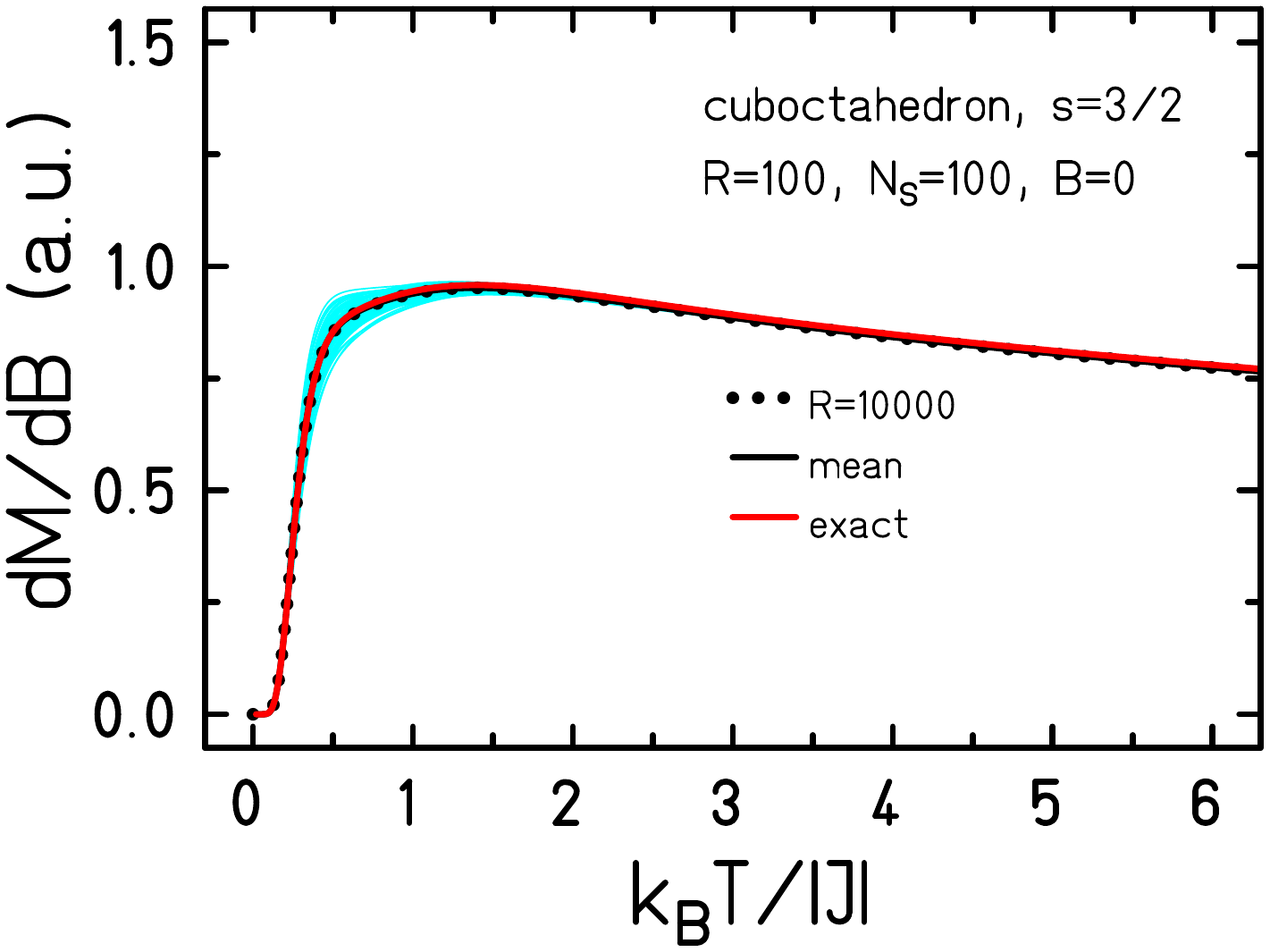}
\includegraphics*[clip,width=0.69\columnwidth]{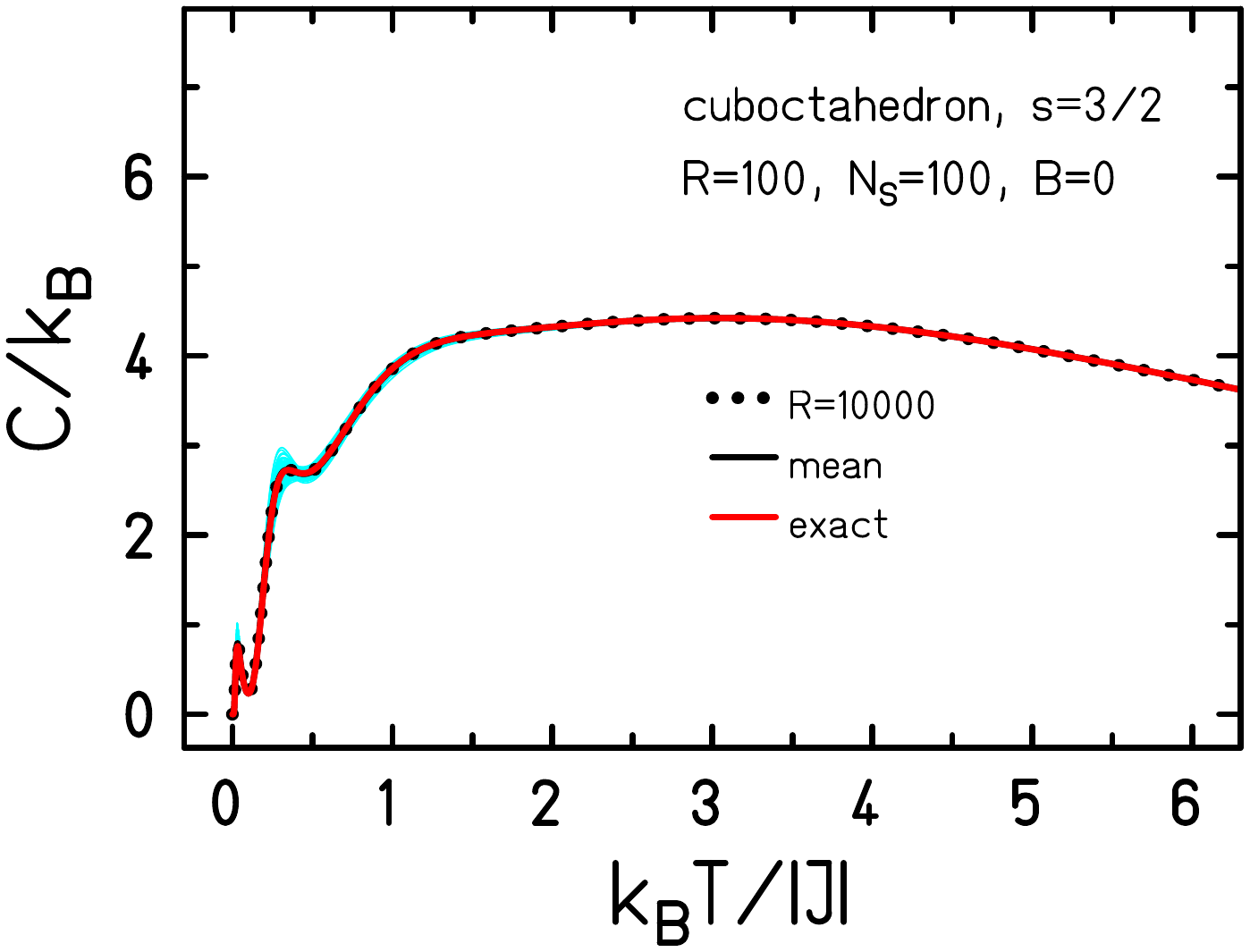}
\caption{Cuboctahedron $N=12$, $s=3/2$: The
  light-blue curves depict 100 different estimates ($R=100$) of the
  differential susceptibility as well as the heat capacity.
  Mean values as well as the exact result are also
  presented.}
\label{ftlm-accuracy-f-B2}
\end{figure}

\begin{figure}[ht!]
\centering
\includegraphics*[clip,width=0.69\columnwidth]{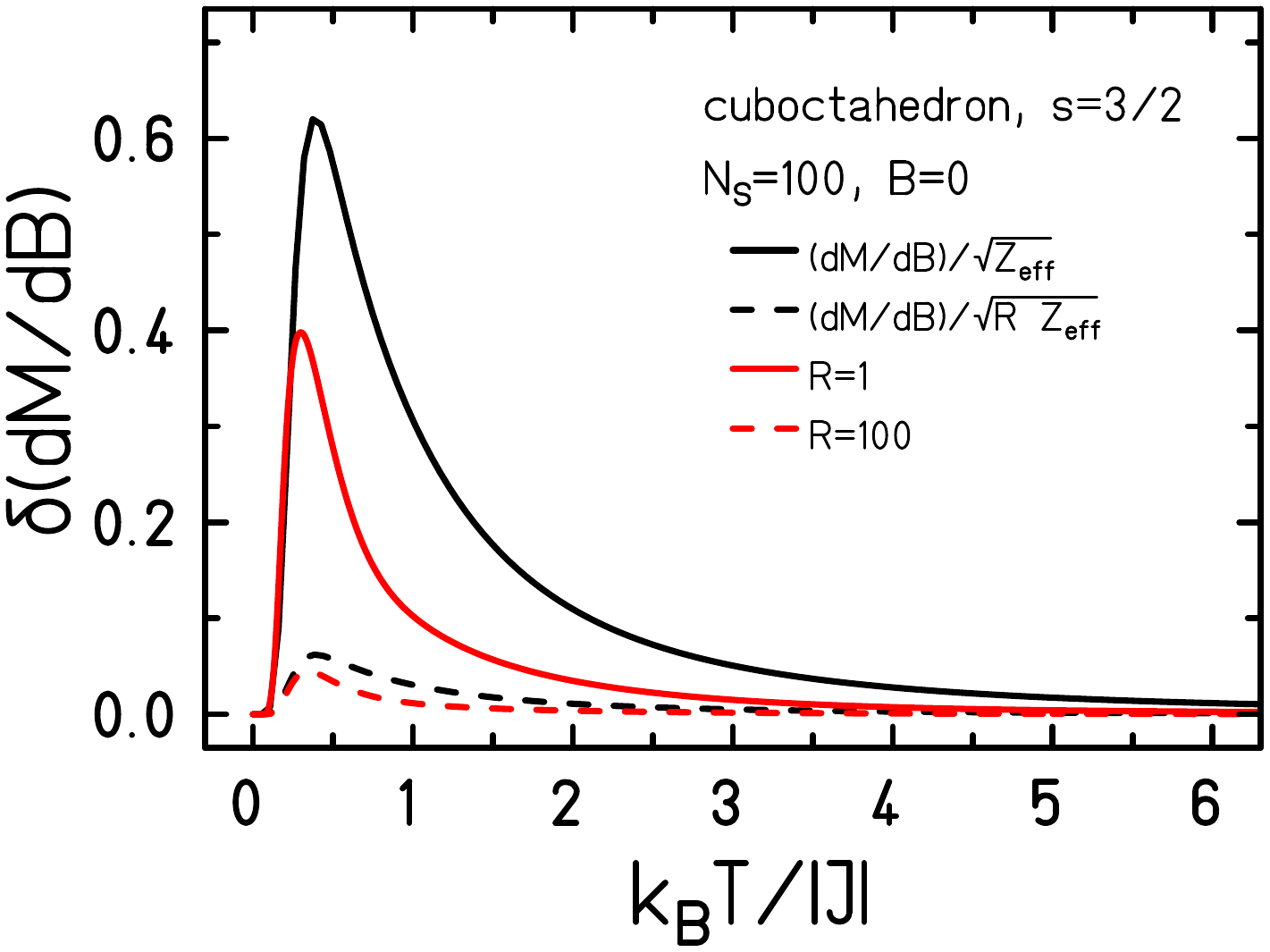}
\includegraphics*[clip,width=0.69\columnwidth]{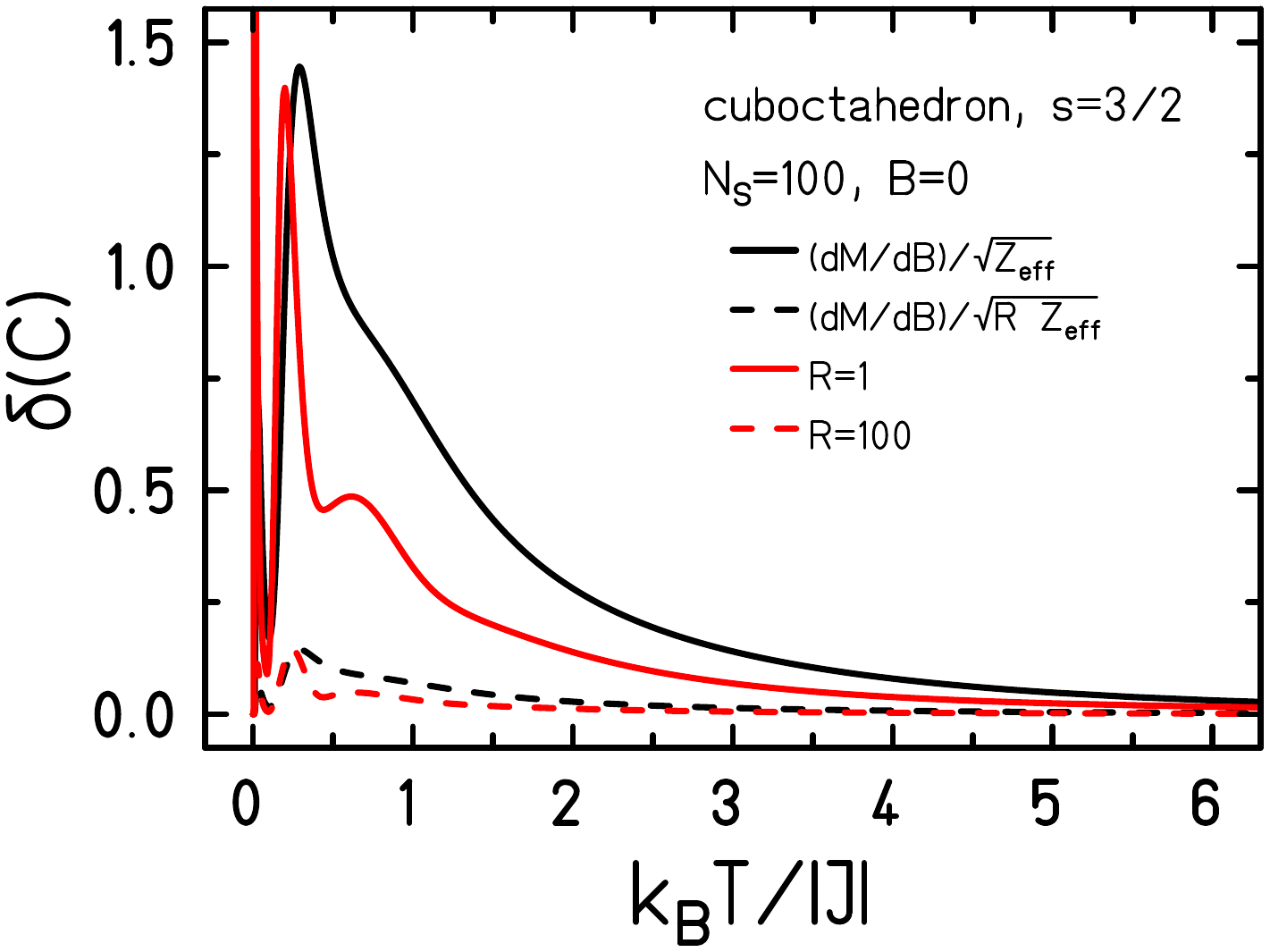}
\caption{Cuboctahedron $N=12$, $s=3/2$: Standard
  deviation of the differential 
  susceptibility as well as the heat capacity compared to the
  error estimate for two sets of estimates, one with $R=1$
  (solid curves) and one with $R=100$ (dashed curves).}
\label{ftlm-accuracy-f-B3}
\end{figure}

Since the system is not too big we repeated this analysis for
$N_S=100$ samples of FTLM estimates with $R=100$ each (in this
case $O^{\text{m}}(T,B)$ of \eqref{E-2-E} equals
$O^{\text{FTLM}}(T,B)$ of \eqref{E-2-C}).
The result is shown in \figref{ftlm-accuracy-f-B2}. One immediately
recognizes the much smaller spread of the estimates. Only at
(sharp) features of the respective functions deviations are
still visible. The origin can be found in strong variations of
the true density of states with energy and/or external magnetic
field; such variations seem to be hard to emulate by the coarse
grained coverage through the trace estimator.
The related
standard deviations are expected to further decrease in a
Monte-Carlo-fashion by a factor of $1/\sqrt{R}$, compare
\cite{PRE:COR17}. This is indeed found as depicted in
\figref{ftlm-accuracy-f-B3}. The solid curves display the true
standard deviation as well as the estimate for $R=1$, whereas
the dashed curves do the same but for $R=100$. Since
$\sqrt{100}=10$, the fluctuations of the trace estimator should
be ten times smaller, which agrees with the numerical study.

\begin{figure}[ht!]
\centering
\includegraphics*[clip,width=0.69\columnwidth]{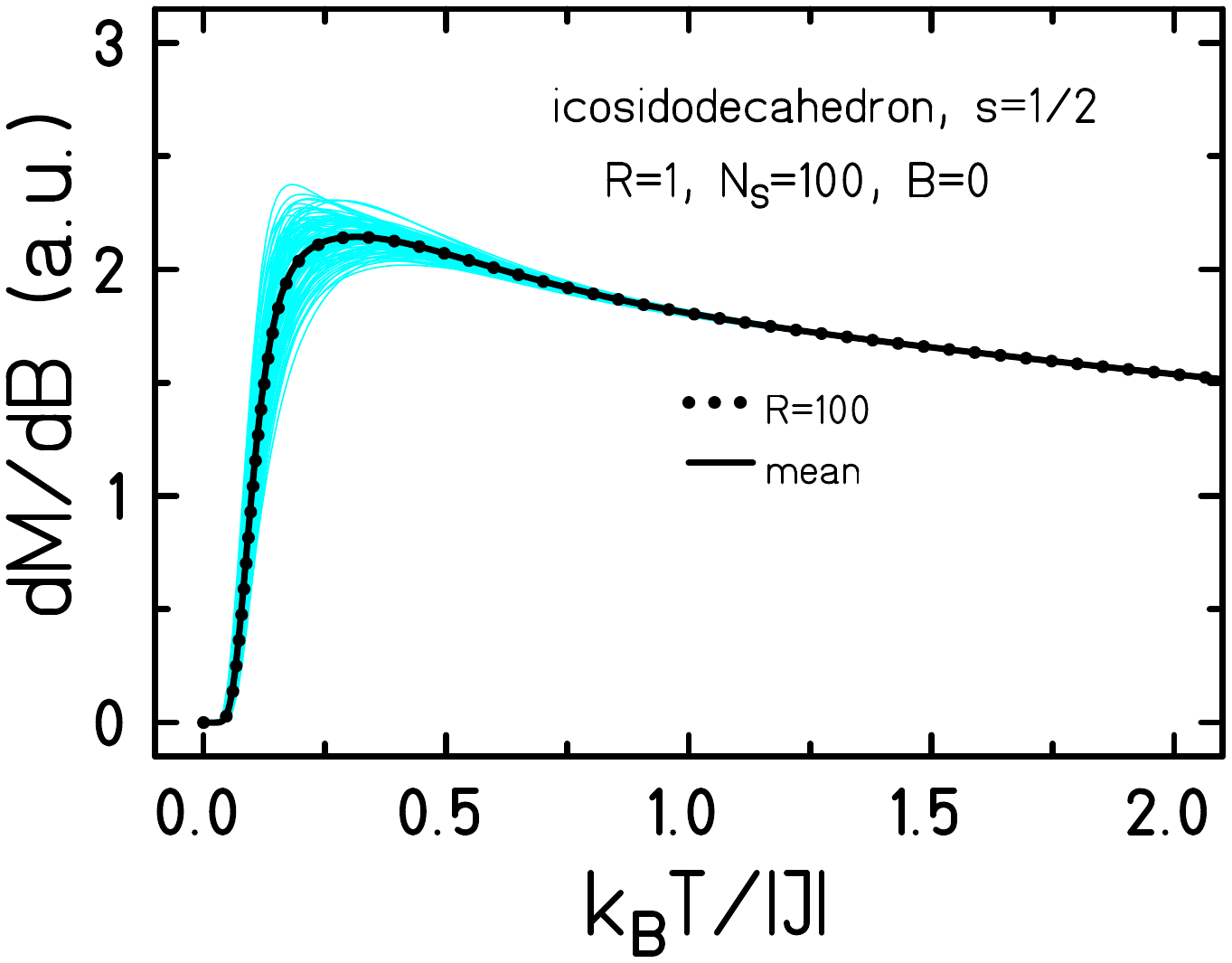}
\includegraphics*[clip,width=0.69\columnwidth]{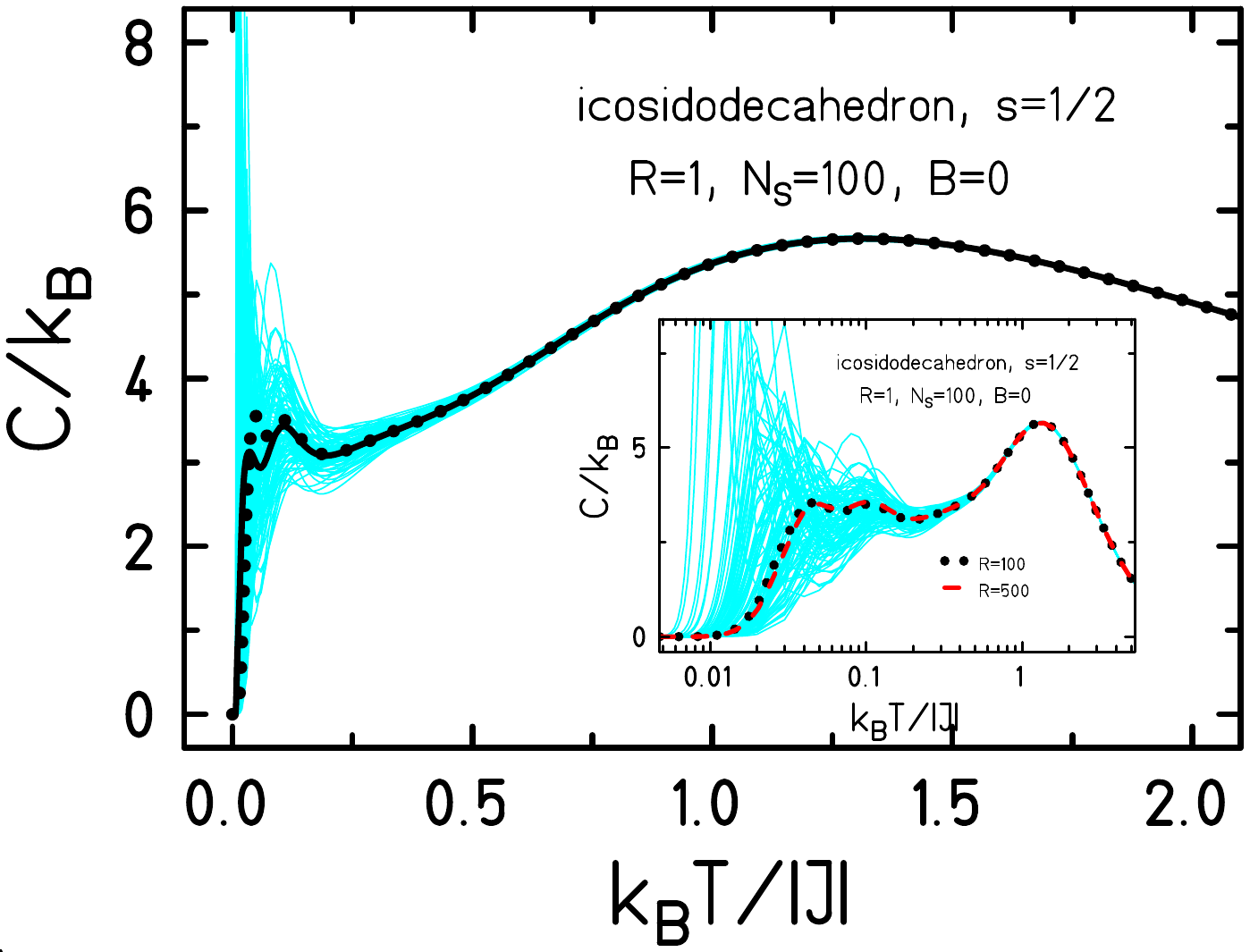}
\caption{Icosidodecahedron $N=30$, $s=1/2$: The
  light-blue curves depict 100 different estimates of the
  differential susceptibility as well as the heat capacity.
  Mean values are also presented.}
\label{ftlm-accuracy-f-C1}
\end{figure}

\begin{figure}[ht!]
\centering
\includegraphics*[clip,width=0.69\columnwidth]{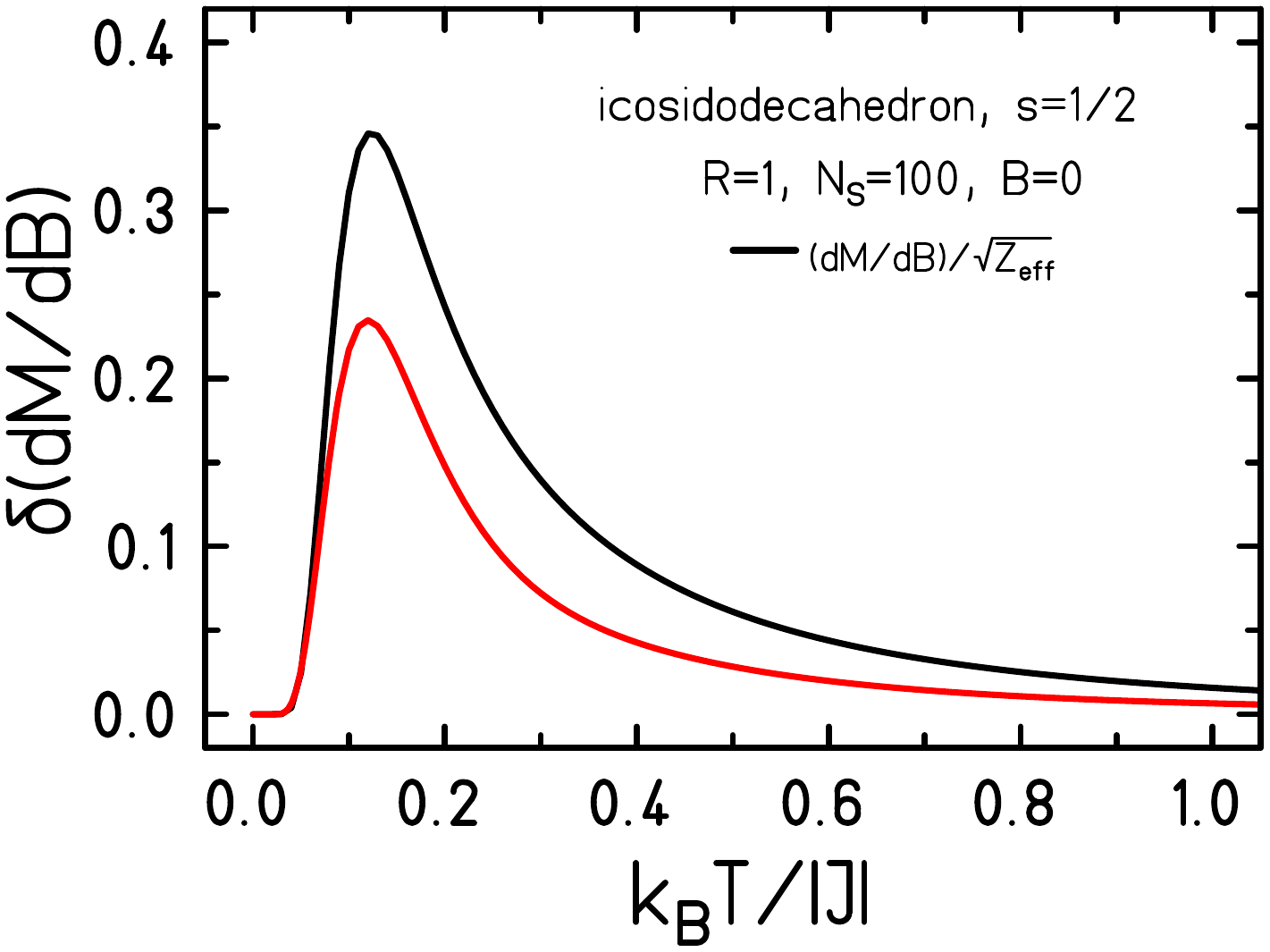}
\includegraphics*[clip,width=0.69\columnwidth]{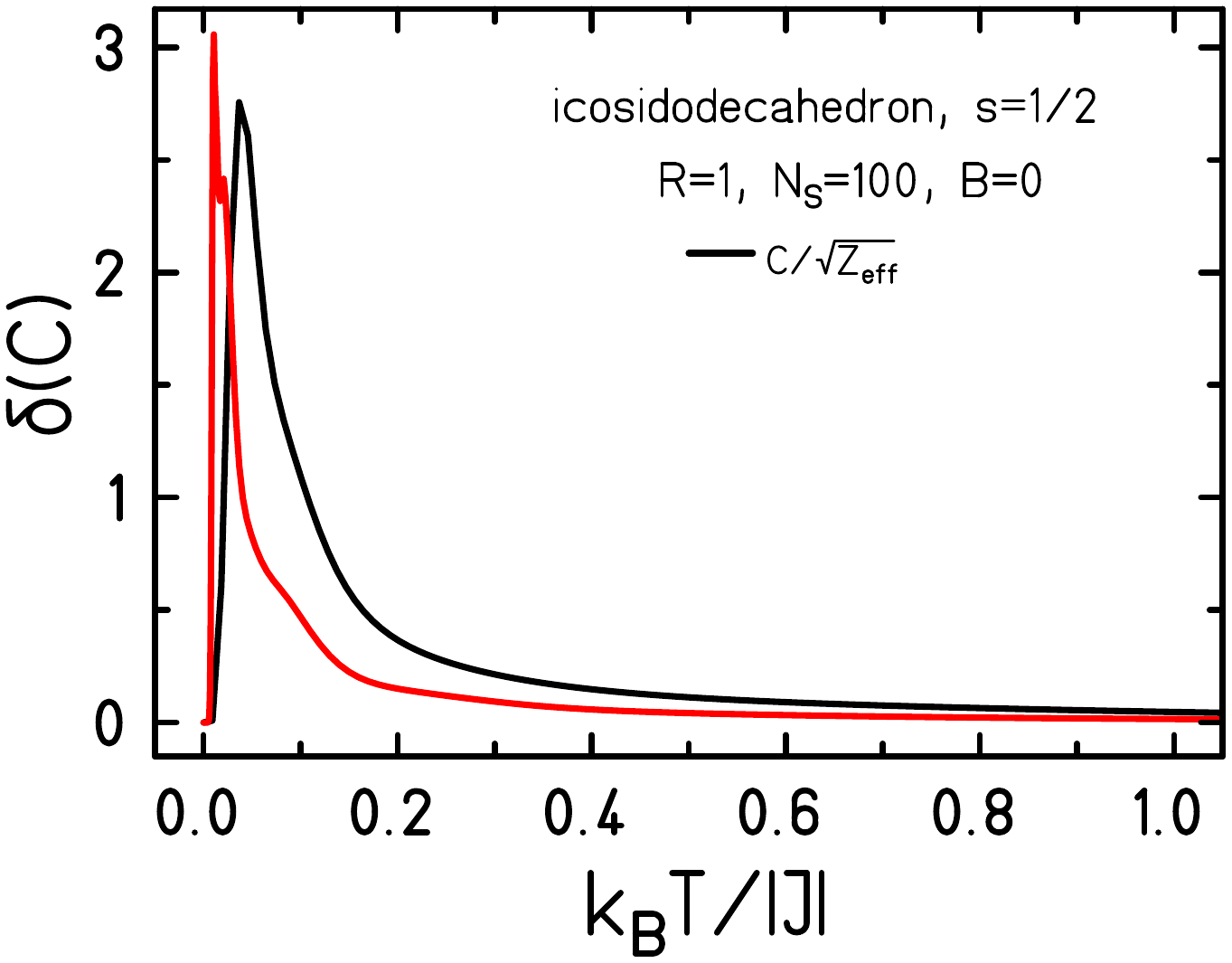}
\caption{Icosidodecahedron $N=30$, $s=1/2$: Standard
  deviation (red) of the differential 
  susceptibility as well as the heat capacity compared to the
  error estimate (black).}
\label{ftlm-accuracy-f-C2}
\end{figure}

The icosidodecahedron -- a keplerate molecule -- could be
synthesized with various ions leading to single spin quntum
numbers of $s=1/2$
\cite{BKH:CC05,TMB:CC09}, $s=3/2$ \cite{TMB:ACIE07,SPK:PRB08}, and
$s=5/2$ \cite{MSS:ACIE99,MLS:CPC01}. Only the spin-1/2-version
can be calculated by means of trace estimators (typicallity
methods) since the dimension of the total Hilbert space is
1,073,741,824. For $s=1$ it would already be $\sim 2\cdot
10^{14}$ and thus out of reach for such methods.

The spectrum of the icosidodecahedron features similar
properties as that of the cuboctahedron or that of finite-size
realizations of the \kagome\ lattice
\cite{SSR:JMMM05,RLM:PRB08,SSR:PRB18}: the spectrum is rather
dense, which in particular means that many singlets populate the
energy spectrum below the lowest triplet level. The latter has a
stark impact on the low-temperature behavior of the heat
capacity. On the other hand, for a given temperature a dense
spectrum leads to a much 
larger effective partition function $Z_{\text{eff}}$ in
\fmref{E-1-3} compared to e.g.\ a bipartite spin system with
pronounced low-lying energy gaps and thus to smaller
fluctuations at this temperature. Comparing
Figs.~\xref{ftlm-accuracy-f-A1}
and \xref{ftlm-accuracy-f-C1}, one notices that the fluctuations
of the estimators were visible below $k_B T\approx 10 |J|$ for
the ferric wheel,
whereas this value is only $k_B T\approx 1 |J|$ for the
icosidodecahedron. This means, that a single random vector is
sufficient for the evaluation of these observables $k_B T\gtrapprox 1 |J|$, which
constitutes a drastic reduction of the computational effort. 
One could thus sloppily say that frustration
works in favor of trace estimators, cf. Ref.~\cite{PrK:PRB18}. 

The respective standard deviations support these
impressions. Only at the lowest temperatures -- corresponding to
the low-lying level structure in particular of the singlet
states -- the specific heat estimates express large
fluctuations. The susceptibility is not affected, since the
low-lying singlets are non-magnetic.

\subsection{Delta chain}
\label{sec-3-3}

The above discussed spin systems have a pretty regular
(dome-shaped, close to Gaussian) density of states. As our next
example we would like to investigate a delta chain (also
sawtooth chain) close to
the quantum critical point \cite{KDN:PRB14,DmK:PRB15}. We choose
a chain of $N=32$ sites with a ferromagnetic nearest-neighbor
interaction $J_1$ and a next-nearest neighbor antiferromagnetic
interaction $J_2$
between spins on adjacent odd sites, i.e. $i$ and $i+2$ with
$i=1,3,5,\dots$. Periodic boundary conditions are applied.
At the quantum critical point (QCP), $|J_2/J_1|=1/2$, the system
features a massive degeneracy due to multi-magnon flat
bands. Therefore, close to the QCP an additional small energy
scale is created, around which the density of states exhibits an
additional low-energy maximum. It is worth mentioning that such
a compound, that is very close to the QCP, could be synthesized
recently \cite{BML:npjQM18}. 

\begin{figure}[h]
\centering
\includegraphics*[clip,width=0.69\columnwidth]{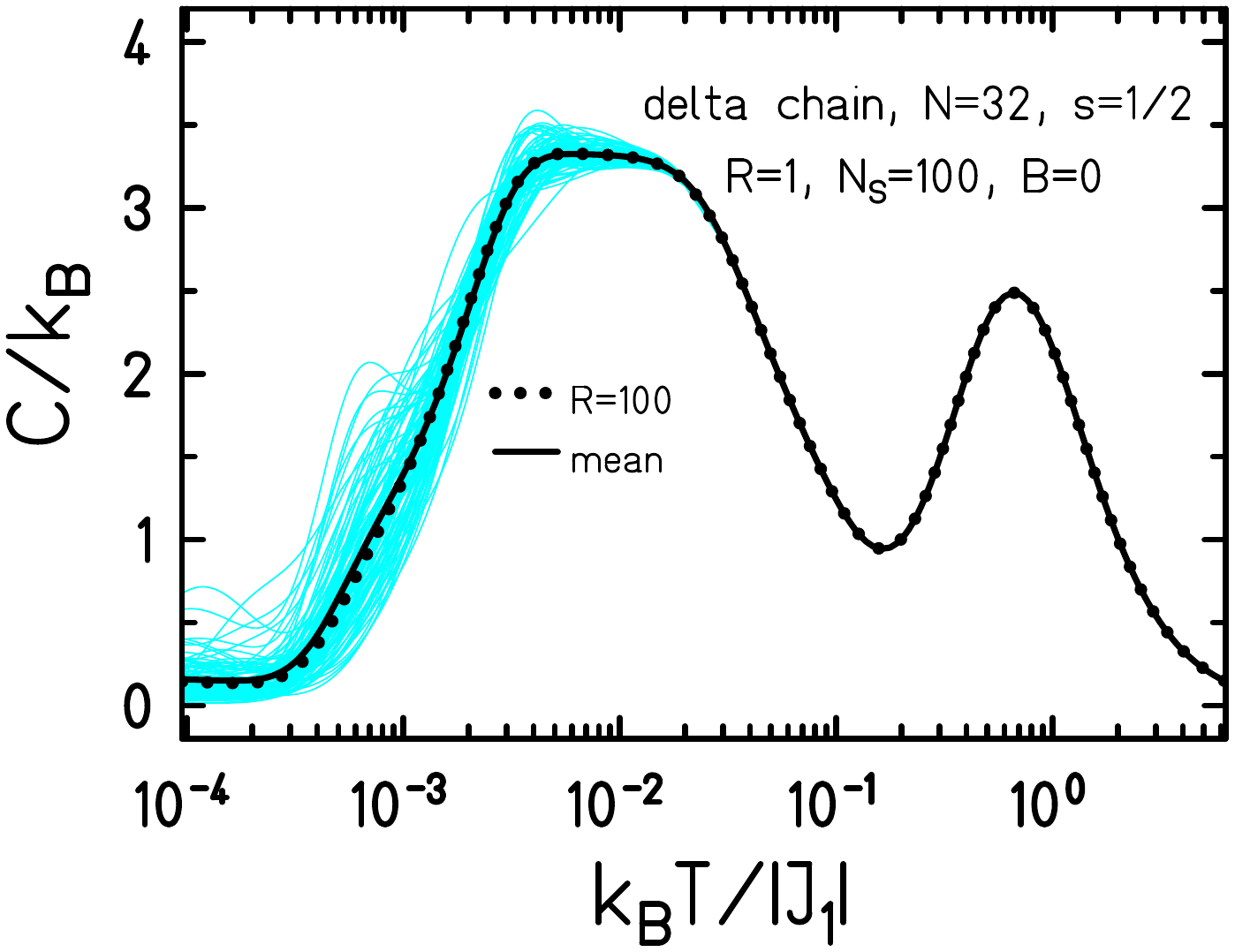}
\includegraphics*[clip,width=0.69\columnwidth]{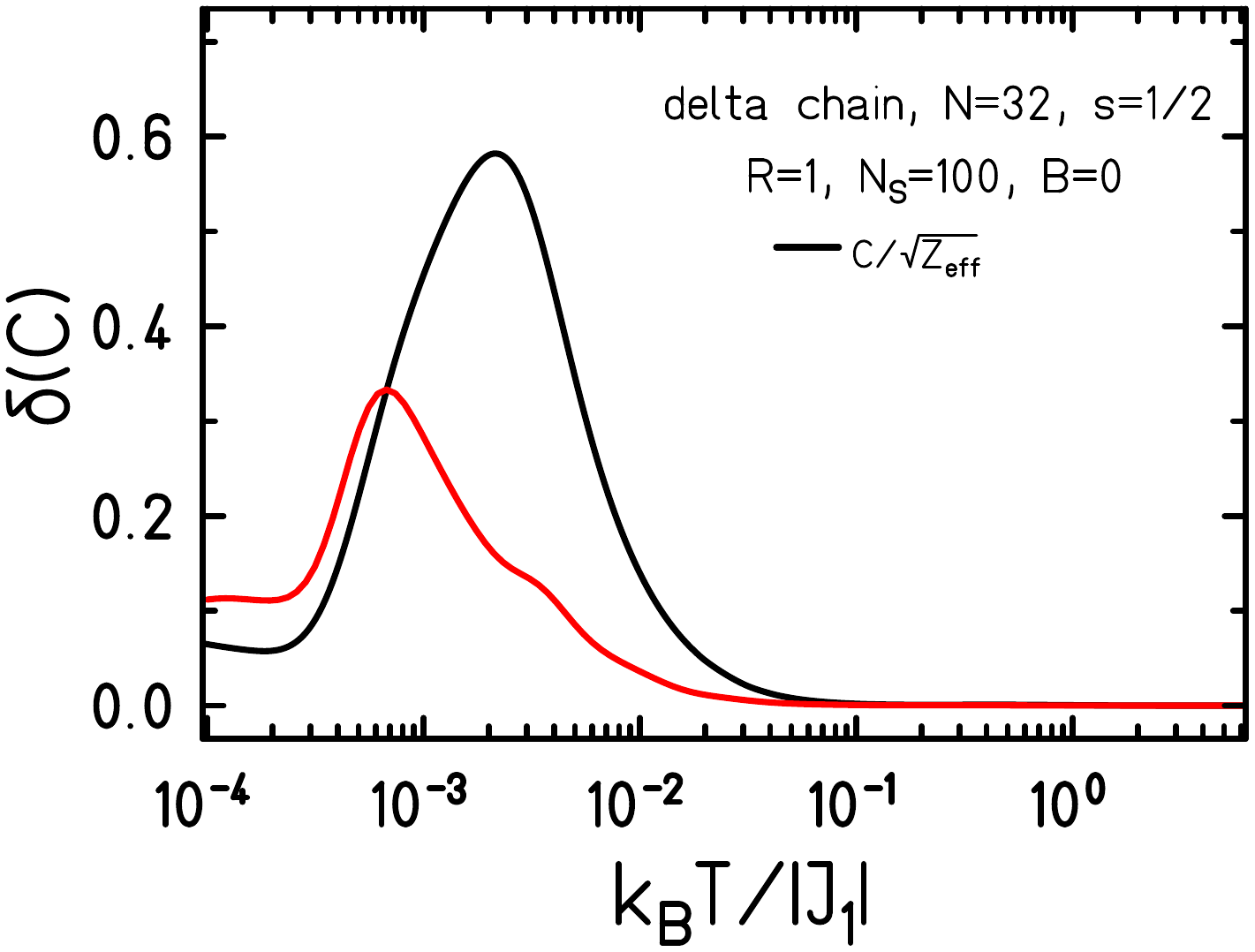}
\includegraphics*[clip,width=0.69\columnwidth]{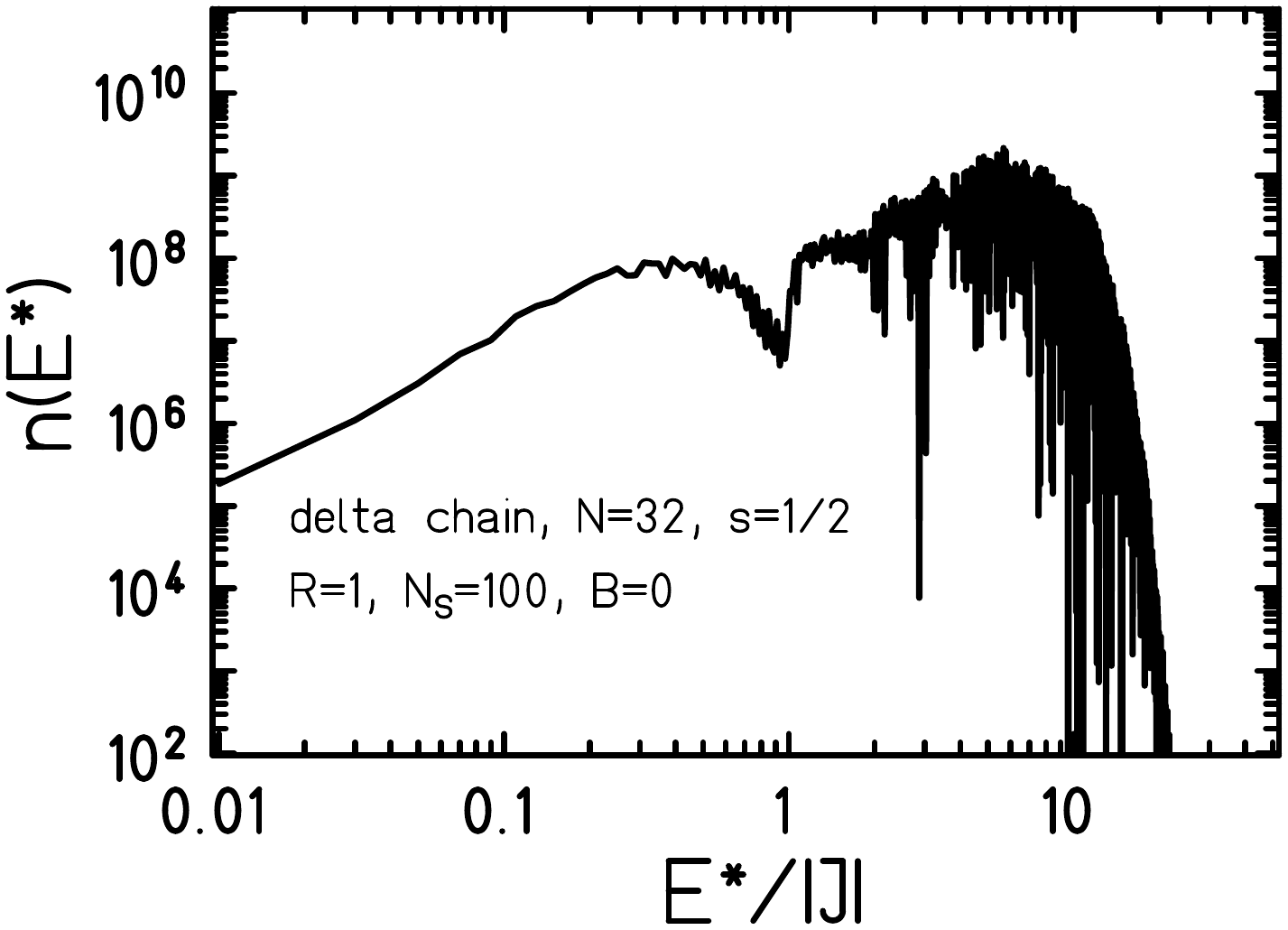}
\caption{Delta chain $N=32$, $s=1/2$, $|J_2/J_1|=0.45$: heat
  capacity, standard deviation and density of states (from top
  to bottom). The light-blue curves depict 100 different estimates of the
  heat capacity. Mean values are also presented. The density of
  states is presented for excited levels with excitation energy
  $E^*$.}
\label{ftlm-accuracy-f-D}
\end{figure}

When evaluating the estimates one notices that fluctuations of
observables appear only for temperatures of the order of the
emergent small energy scale, as can be seen in
\figref{ftlm-accuracy-f-D}. This energy scale, which is
approximately $10^{-2} |J_1|$ and corresponds to the
low-temperature maximum, is much smaller than the dominant
scale $|J_1|$, that corresponds to the high-temperature Shottky
peak. 

The reason for this behavior can be traced back to the enormous
number of low-lying levels assembled at the low-energy scale,
compare density of states in \figref{ftlm-accuracy-f-D},
that at temperatures elevated above the low-temperature scale
contribute to the effective partition function \fmref{E-1-3}
and thus lead to a very small estimate for the fluctuations of
any observable. This is clearly seen for $\delta(C)$ in
\figref{ftlm-accuracy-f-D}, which virtually drops to zero above
the low-temperature scale. Also in this case a single random
vector suffices to evaluate the thermal behavior above the
low-temperature scale.

\subsection{An integrable spin system}
\label{sec-3-4}

We already mentioned that the use of the concept of typicality
for trace estimators is not connected to the question whether
ETH holds for the respective system or not. Here we present a
simple example of a spin-1/2 chain with nearest-neighbor
antiferromagnetic interaction, that
would be integrable via the Bethe ansatz
\cite{Bet:ZP31,KRS:LMP81,Tak:PLA82,Bab:NPB83}. We investigate a
spin chain of $N=24$ spins $s=1/2$ with periodic boundary
conditions that can also be solved numerically exactly when
employing the symmetry groups SU(2) and C$_N$
\cite{HeS:PRB19}. For the FTLM investigation a reduced symmetry
was used, namely $\op{S^z}$--symmetry as well as translational
symmetry C$_N$. 

\begin{figure}[ht!]
\centering
\includegraphics*[clip,width=0.69\columnwidth]{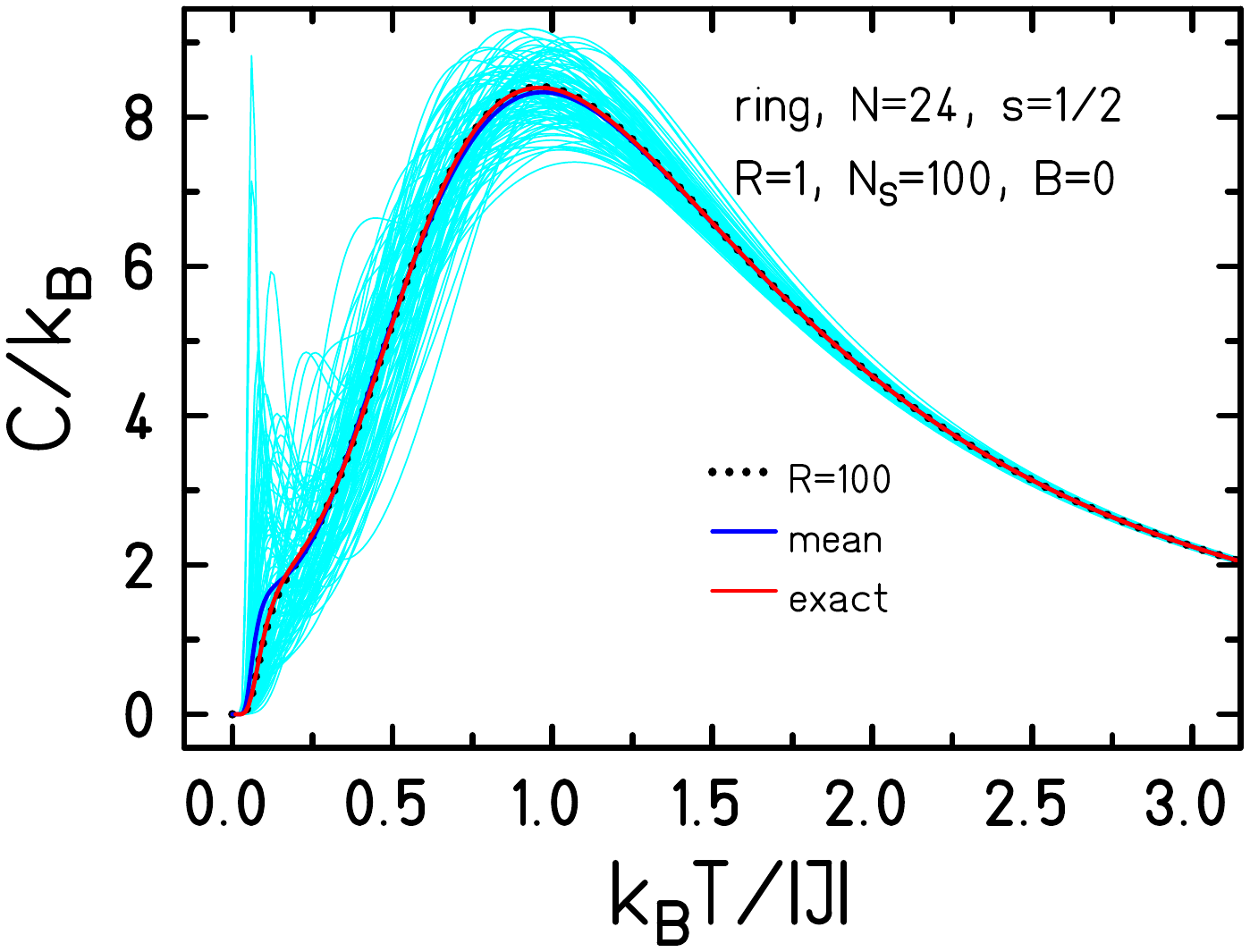}
\caption{Spin ring $N=24$, $s=1/2$: The
  light-blue curves depict 100 different estimates of the heat capacity.
  Mean values as well as the exact result are also presented.}
\label{ftlm-accuracy-f-E}
\end{figure}

\begin{figure}[ht!]
\centering
\includegraphics*[clip,width=0.69\columnwidth]{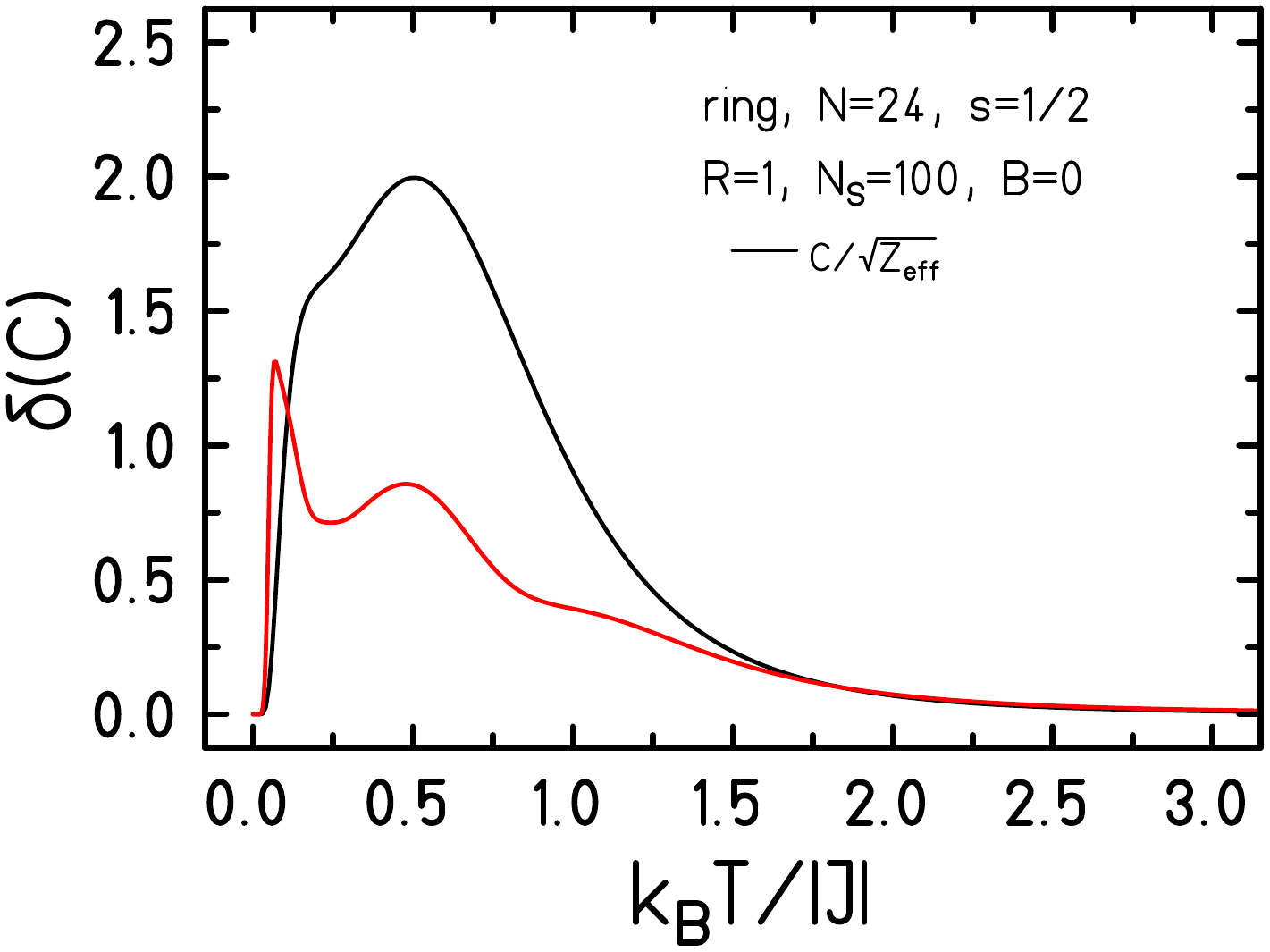}
\caption{Spin ring $N=24$, $s=1/2$: Standard deviation (red) of the
  heat capacity compared to the error estimate (black).}
\label{ftlm-accuracy-f-F}
\end{figure}

In Figs.~\xref{ftlm-accuracy-f-E} and \xref{ftlm-accuracy-f-F}
we present our results for the heat capacity. The susceptibilty
(not shown) behaves similarly. The results are very similar to
the already discussed examples. The largest deviations again
occur at and below a temperature scale of the order of the
exchange interaction $|J|$. A sharp peak of the standard
deviation $\delta(C)$ at very low temperatures occurs at
temperatures coressponding to the lowest (singlet-triplet) gap.
However, one would expect for this class of spin systems
that the temperature above which the approximation is good drops
with increasing system size since the lowest gaps shrink for
this system that is gapless in the thermodynamic limit.

\subsection{A Haldane  spin system}
\label{sec-3-5}

\begin{figure}[ht!]
\centering
\includegraphics*[clip,width=0.69\columnwidth]{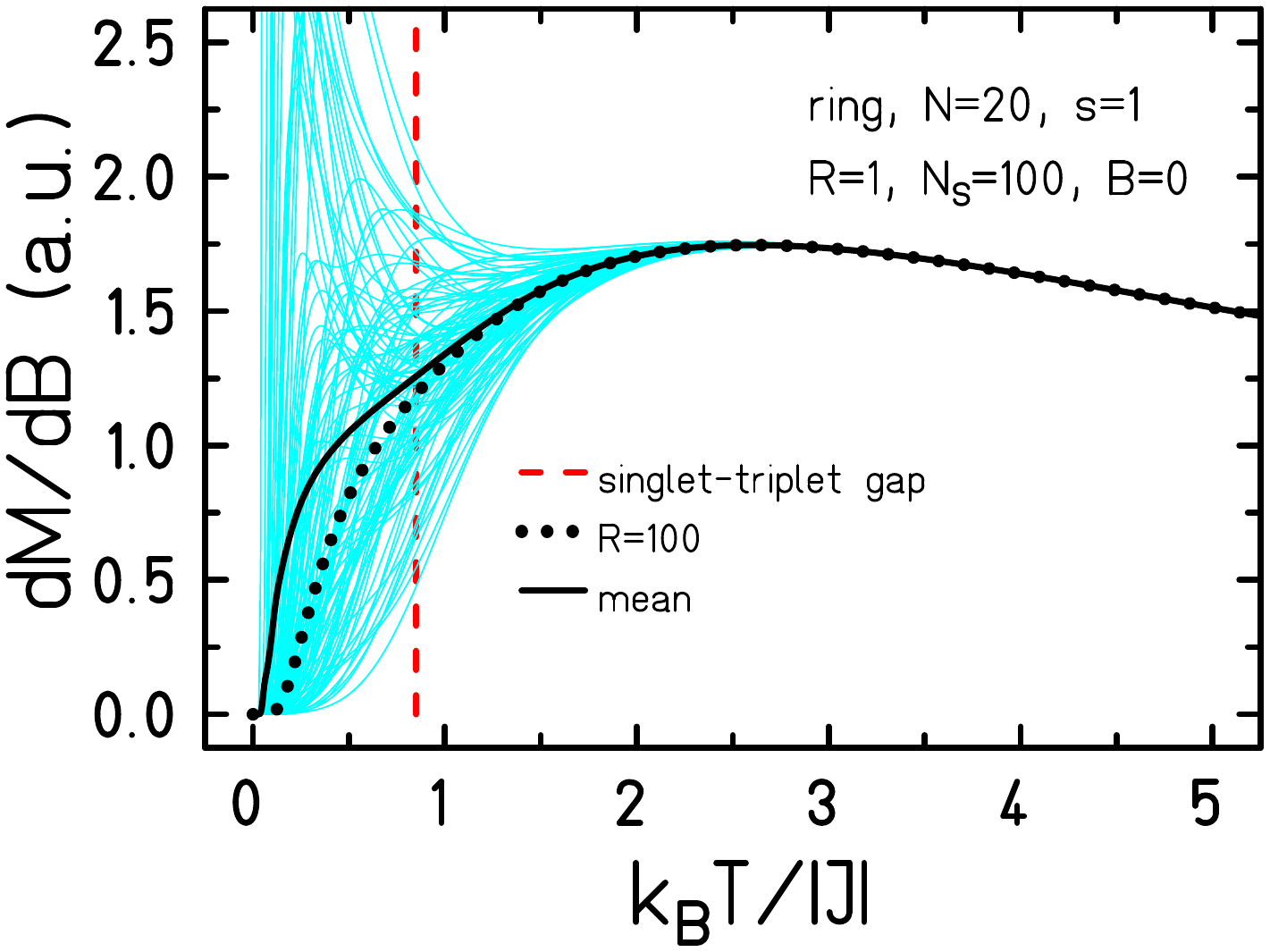}
\includegraphics*[clip,width=0.69\columnwidth]{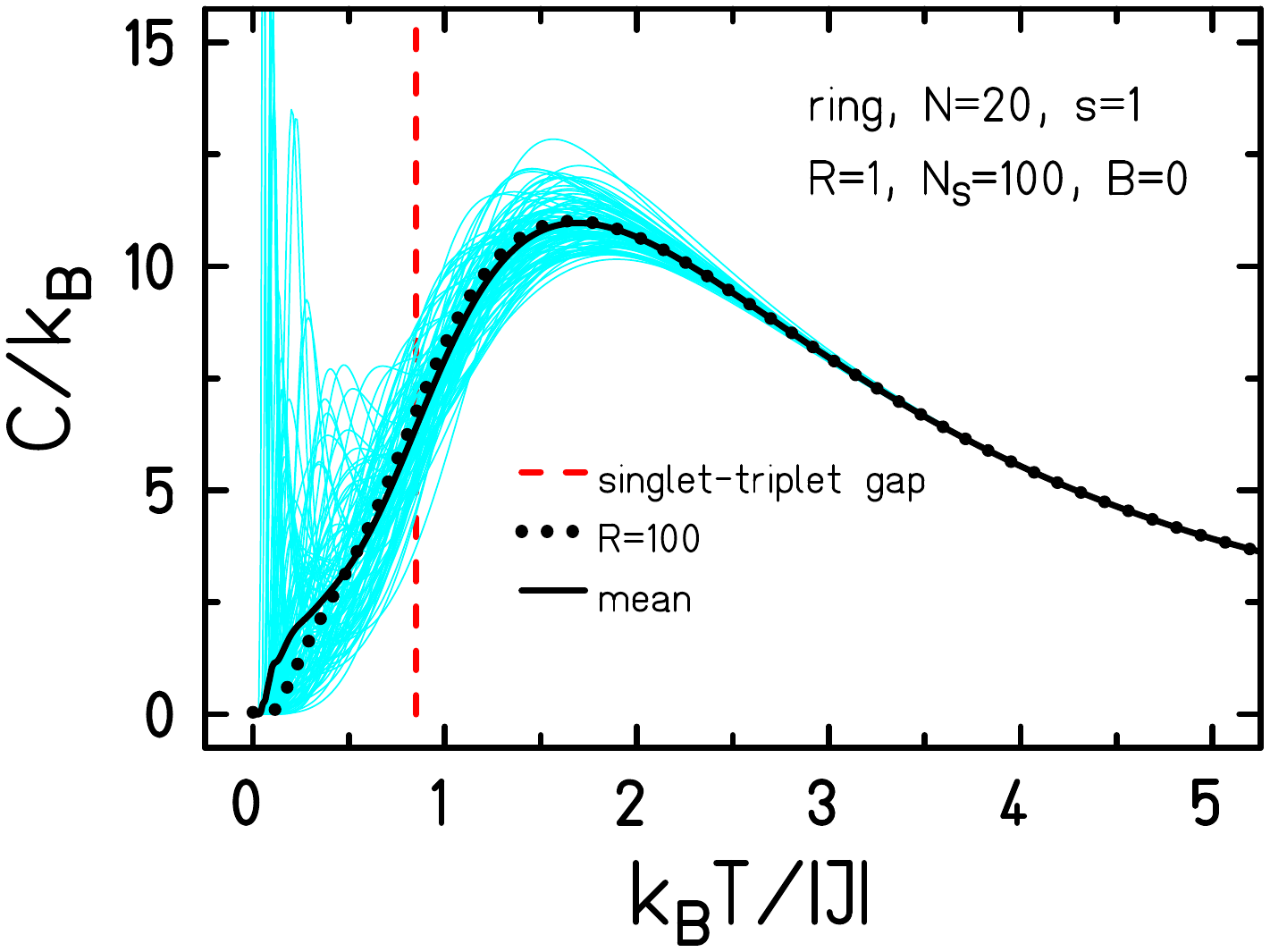}
\caption{Spin ring $N=20$, $s=1$: The
  light-blue curves depict 100 different estimates of the
  differential susceptibility as well as the heat capacity.
  Mean values as well as the exact result are also
  presented.}
\label{ftlm-accuracy-f-G}
\end{figure}

\begin{figure}[ht!]
\centering
\includegraphics*[clip,width=0.69\columnwidth]{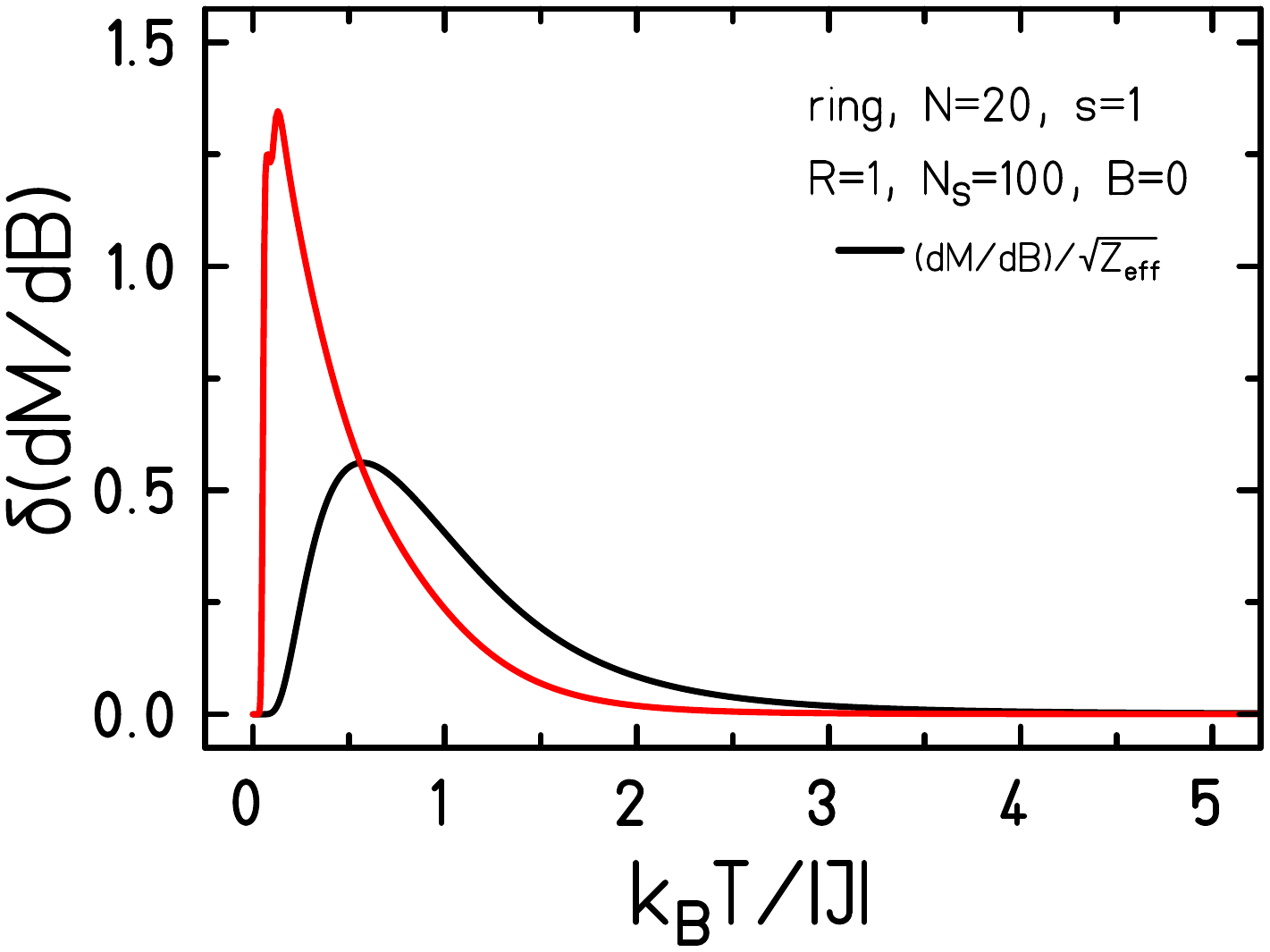}
\includegraphics*[clip,width=0.69\columnwidth]{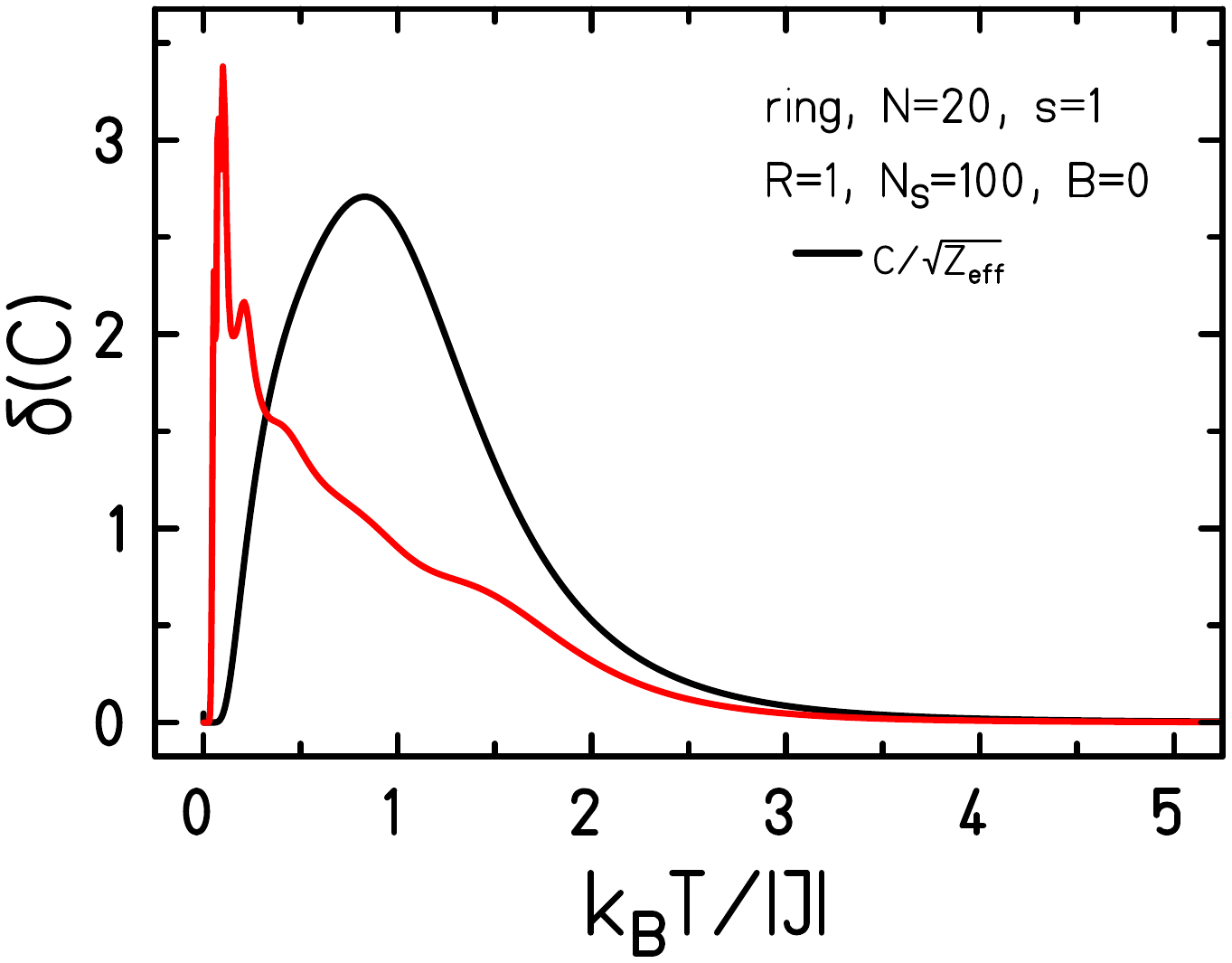}
\caption{Spin ring $N=20$, $s=1$: Standard deviation (red) of the differential 
  susceptibility as well as the heat capacity compared to the
  error estimate (black).}
\label{ftlm-accuracy-f-H}
\end{figure}

The question how the lowest gap influences the low temperature
quality of the approximation will be addressed in this
section. To this end we choose a Haldane spin chain of $N=20$
and $s=1$ with nearest neighbor antiferromagnetic exchange as an
example for systems where the lowest gap is sizable and does not
even close in the thermodynamic limit.

As one can see in \figref{ftlm-accuracy-f-G} both susceptibility
as well as heat capacity fluctuate largely about and below a
temperature  scale that is provided by the lowest energy gap
(dashed vertical line). In essence this means -- here and for
all previous examples -- that a single random vector,
\eqref{E-2-A}, does not work for temperatures of this order and
below, except for $T=0$ where the method is bound to be
exact \cite{JaP:AP00,PRE:COR17}. This is also clearly reflected
by the respective standard deviations shown in
\figref{ftlm-accuracy-f-H}. In addition, here we 
encounter an example where the standard deviation
$\delta(\dint\mathcal{M}/\dint B)$ assumes values much larger
than our estimator \fmref{E-1-3} for $T>0$. 

\begin{figure}[ht!]
\centering
\includegraphics*[clip,width=0.69\columnwidth]{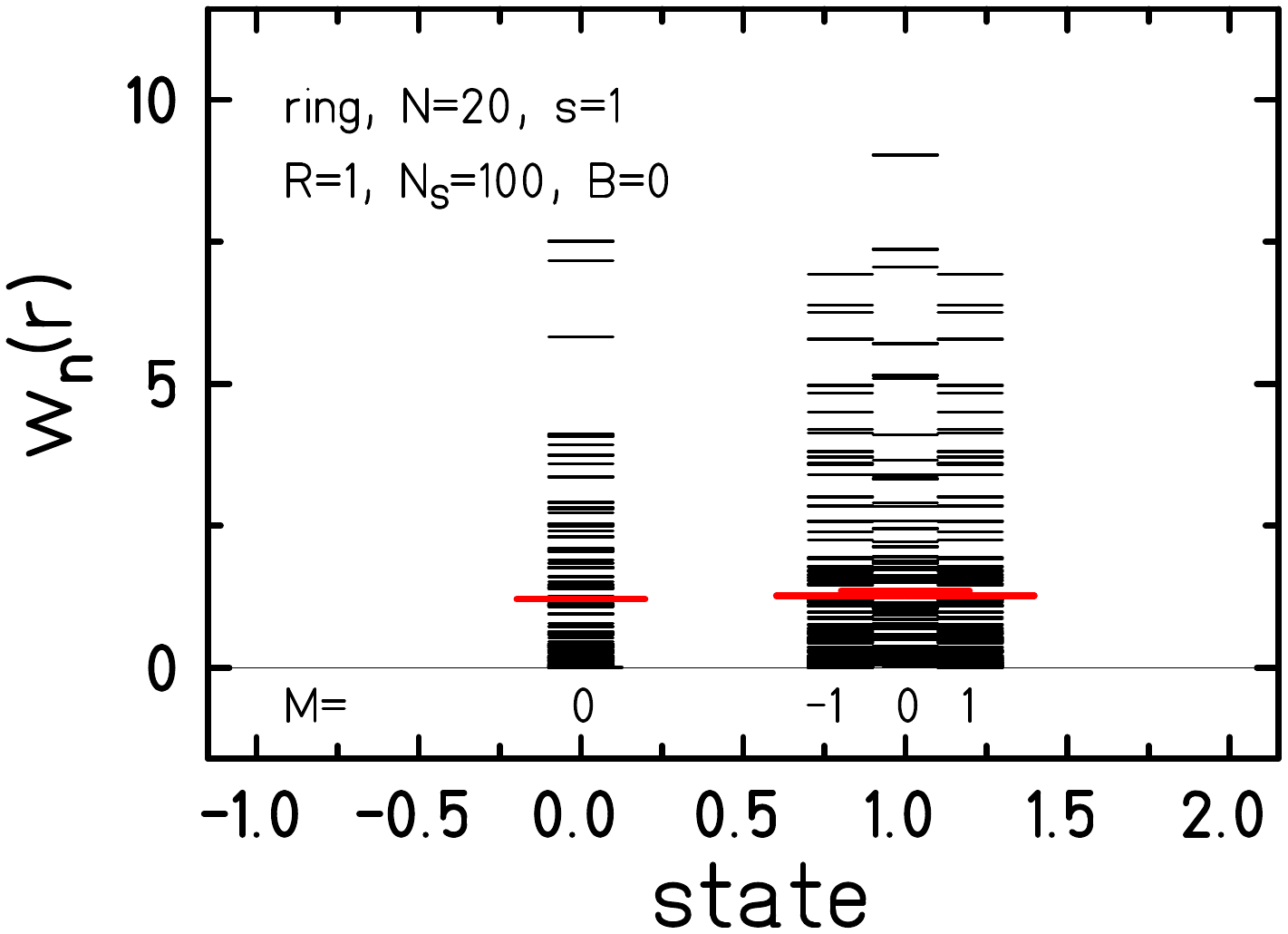}
\caption{Spin ring $N=20$, $s=1$: Weights $w_n(r)$ of the ground
  state singlet and the first excited triplet (black bars). Mean
  weigths are depicted by thick red bars. Technically, the
  weights for $M=-1$ are a simple copy of those for $M=+1$ thanks
  to symmetry.}
\label{ftlm-accuracy-f-I}
\end{figure}

The reason of the strong fluctuations of low-temperature
observables lies in a poor coverage of $w_n(r)$ for the lowest
energy eigenvalues. Although the lowest eigenvalues, thanks to
the properties of power methods, are very accurate, 
i.e. possess a standard deviation of $< 10^{-4}$ or better, the
corresponding weight factors fluctuate largely. This can be seen
in \figref{ftlm-accuracy-f-I} where the weigths $w_n(r)$ are
displayed for the ground state and the first excited state as
they are evaluated in the respective Hilbert subspaces
\subhm. These fluctuations of the weights are conserved by a power method,
this is particularily important for the low-lying levels.
In order to yield an accurate low-temperature partition function
the weights of the lowest states should equal one in \fmref{E-2-D}.
The naturally occuring variation of the weights in a random vector
are amplified at low temperature by the Boltzmann factor.
One can derive an estimate for the relative error of the specific
heat assuming (for simplicity) that at low temperatures only the
ground state as well as the first excited state contribute to
the partition function,
\begin{eqnarray}
\label{E-3-F}
\frac{\delta(C)}{C}
&\approx&
\left|
\frac{\frac{w_1}{w_0}
  (1+d\exp[-\beta\Delta])^2}
{(1+\frac{w_1}{w_0} d \exp[-\beta\Delta])^2}
-1
\right|
\\
\label{E-3-FB}
&\rightarrow&
\left|
\frac{w_1}{w_0}
-1
\right|
\ , \ \text{for}\ \beta\rightarrow\infty
\ .
\end{eqnarray}
Here $\Delta$ is the gap between ground and first excited state,
$d$ the degeneracy of the first excited state, and 
$w_0$ and $w_1$ the weights of the ground and the first excited state.
The ratio of both dominates the relative error at low temperatures;
it can assume large values.
This problem is independent of the observable at hand,
it stems from the fluctuations that are intrinsic to the random vector.
Therefore, \eqref{E-3-FB} does not only hold for the heat capacity,
which is a variance
$\left\langle\op{H}^2\right\rangle
-
\left\langle\op{H}\right\rangle^2$, but for
$\left\langle\op{H}^2\right\rangle$
and
$\left\langle\op{H}\right\rangle$
separately.

Although every random vector possesses these fluctuations,
a proper averaging as outlined in \fmref{E-2-D} decreases the
fluctuations drastically. Figure~\xref{ftlm-accuracy-f-I} demonstrates that
after averaging over 100 random vectors, the averaged weights
(thick red bars) approach an equal magnitude although not yet
one, but $\sim 1.2$. For most observables it is sufficient that
the averaged weights of the lowest states are about the same
since they appear simultaneously in numerator and denominator of
\eqref{E-2-C}. Quite recently new ideas have been formulated how
to improve the low-temperature estimates of FTLM also for small
numbers $R$ of random vectors by taking special care of the
weights for low-lying levels \cite{MoT:A19}.

The investigation of the averaged weights of low-lying levels
also sheds light on the failure of the naive mean according to
\eqref{E-2-B}. This kind of a mean does not average the
individual weights, but the individual single-vector expectation
values, which converges either very slowly or even to a
different function at low temperatures.

\section{Discussion and conclusions}
\label{sec-4}

Finally, we can conclude that typicality-based methods allow an
astonishingly accurate approximation of static thermodynamics
observables sometimes using just one random vector. This
qualifies methods such as FTLM for a reliable 
treatment of large quantum systems, in particular of those that
cannot be delt with by quantum Monte Carlo due to the sign
prolem and of those where approximations using matrix product
states converge slowly such as the two-dimensional
\kagome\ lattice \cite{SSR:PRB18}.

The simple idea of typicality to replace a trace by an
expectation value with respect to just one random vector works
indeed for large enough temperatures.
$1/\sqrt{Z_{\text{eff}}}$
provides the estimate for the relative error to be expected
for temperatures well above the lowest excitation gap.
An additional average over many random vectors according to
\fmref{E-2-C} further increases
the accurcy in a Monte-Carlo fashion and reduces the error by
another factor $1/\sqrt{R}$, where $R$ is the number of employed
random vectors. The simple average \fmref{E-2-B} of
single-vector approximations does not converge properly
especially at low temperatures.

Although power methods such as the Lanczos method
yield exact ground state expectation values
for systems with non-degenerate ground states, and should thus
be accurate at $T=0$, the large fluctuations of estimates using a
single random vector surprise.
We could clarify the latter problem by elucidating the important
role jointly played by the 
energy gap between ground state and first excited
state as well as the weight factors of both states.
Although both energies are spectroscopically accurate it
needs sufficient averaging to tame the strongly fluctuating
weight factors. In view of this, and with
$1/\sqrt{Z_{\text{eff}}}$ in mind, one can state that the
discussed approximations work better for systems with small gap
and larger density of low-lying states. Therefore, 
frustration works in favor of trace estimators.

Overall, we conclude that methods such as FTLM, which rely on
trace estimators, are astonishingly accurate. We could demonstrate
with several prototypical examples that the standard deviations
of observables can be systematically reduced via averaging. In
addition, we are convinced that we could provide a valuable
contribution in order to trust these methods by presenting
realistic standard deviations \cite{PhysRevA.83.040001}.

\section*{Acknowledgment}

This work was supported by the Deutsche Forschungsgemeinschaft DFG
(314331397 (SCHN 615/23-1); 397067869 (STE 2243/3-1); 355031190 
(FOR~2692); 397300368 (SCHN~615/25-1)). 
Computing time at the Leibniz Center in Garching is gratefully
acknowledged. J.S. is indebted to Peter Reimann for all his deep
questions that stimulated part of this research. R.S. thanks
Peter Prelov\ifmmode \check{s}\else \v{s}\fi{}ek very much for
fruitful discussions over many years. All authors thank Hans De
Raedt, Peter Prelov\ifmmode \check{s}\else \v{s}\fi{}ek, Patrick
Vorndamme, Katsuhiro Morita as well as George Bertsch for valuable comments.


%

\end{document}